\pdfoutput=1

\documentclass[12pt,preprint]{aastex}

\slugcomment{Accepted for Publication in the Astrophysical Journal}

\shorttitle{UV Spectroscopy of HD 44179}
\shortauthors{Sitko et al.}

\begin{document}

\title{Ultraviolet Spectroscopy of HD 44179}

\author{Michael L. Sitko}
\affil{Dept. of Physics, University of Cincinnati, Cincinnati, OH 45221\\and\\Space Science Institute, Boulder, CO 80301, USA}
\email{sitkoml@ucmail.uc.edu, sitko@spacescience.org}

\author{Lawrence S. Bernstein}
\affil{Spectral Sciences, Inc., Burlington, MA 010803, USA}
\email{larry@spectral.com}

\and

\author{Robert J. Glinski}
\affil{Dept. of Chemistry, Tennessee Technical University, Cookeville, TN 38505, USA}
\email{rglinski@tntech.edu}

\begin{abstract}
We have re-analyzed the ultraviolet spectrum of HD 44179, the central star(s) of the Red Rectangle nebula, providing improved estimates of the column density, rotational, and vibrational temperatures of the 4th Positive A-X system of CO in absorption. The flux shortward of 2200 \AA \  is a complex blend of CO features with no discernible stellar photosphere, making the identification of other molecular species difficult, and the direct derivation of the dust extinction curve impossible. We confirm that the spin-forbidden CO (a-X) Cameron bands are likely produced by either collisional excitation or a chemical reaction, not photoexcitation, but with a higher internal vibrational excitation than previously determined. We also detect the spin-forbidden CO a$^{\prime}$-X, d-X, and e-X absorption features. The hot CO (A-X) bands exhibit a blue-shift of $\sim$300 km s$^{-1}$, likely occurring close to the white dwarf star(s) suspected as the original source of the ultraviolet flux in the system, and forming the base of the outflow of material in the Red Rectangle. The OH ``comet-band'' system near 3000 \AA \ is also analyzed, and estimates of its rovibrational temperatures determined. The source of the molecules studied in this system is still unknown, but may be a combination of gaseous material associated with the star(s), or processed material from the surrounding dust torus. 
\end{abstract}


\section{Introduction}

HD 44179 is a binary star system that is deeply embedded within a dusty torus at the heart of the Red Rectangle nebula. In addition to the ubiquitous organic infrared emission bands (generally thought, but not yet proven, to be polycyclic aromatic hydrocarbons - PAHs), the nebula possesses a broad emission band centered near 6400 \AA \ and a series of narrower bands of molecular origin throughout much of the visible part of the spectrum \citep{cohen75,schmidt80}. Despite a quarter century of investigation, the origin of both the broad and narrow features remain unidentified. 

The star system itself is a single-line spectroscopic binary with a highly eccentric orbit (e$\sim0.38$) with an orbital period of $\sim$300 days, and the masses of both components are sub-solar \citep{vw95,wae96,waters98,men02}. The visible star seems to be a post-asymptotic giant branch (post-AGB) spectral class F star with an ultraviolet (UV) excess. The nature of the secondary star is unknown. \citet{wae96} suggested that it was a low-mass main sequence star, while \citet{men02} favored a high-luminosity white dwarf. This latter model is based largely on the presence of an ultraviolet (UV) excess above that of the post-AGB primary, which was attributed to a white dwarf secondary. In fact, we will show that the UV flux cannot be entirely photospheric.  

The UV spectrum of HD 44179 contains key information about the excitation of the Red Rectangle nebula and the molecular chemistry required to produce the species responsible for its complex spectrum at longer wavelengths. Since their original discovery by \citet{cohen75} and \citet{schmidt80}, both the broad red emission ``hump'', now referred to as the \textit{Extended Red Emission (ERE)}, and the narrower emission bands at visible wavelengths have defied any positive identification. In a study of NGC 7023, another nebula with significant ERE emission, \citet{witt06} concluded that the carrier of the ERE is made from the ionization of precursor species with photons having energies greater than 10.5 eV \ ($\lambda < 1180$ \AA). Thus, information about molecular species bathed by ultraviolet photons in the vicinity of HD 44179 is central to understanding the chemical development of the carrier of the ERE and other emission features.   

The first clues to the nature of molecular interaction with the UV flux occurred with the demonstration that the UV spectrum of HD 44179 is dominated by molecular emission and absorption by CO \citep{sitko83}. All of the identified bands were due to electronic transitions from the A$^{1}\Pi$-X$^{1}\Sigma^{+}$ 4th Positive (4P) system of this molecule. Absorption features at wavelength shortward of 1600 \AA \  from the vibrational states (v$^{\prime}$,v$^{\prime\prime}$) = (0,0) to at least (8,0) were visible (where v$^{\prime}$ and v$^{\prime\prime}$ refer to the vibrational levels of the upper and lower electronic states, respectively). The entire emission band system seen at  wavelengths longer than 1600 \AA \ was  then due to downward fluorescent emission from the upper (A) electronic state to the ground (X) electronic state. Based on the  strengths of the (0,0) and (1,0) absorption bands, a column density N$_{CO}\gtrsim10^{18}$ cm$^{-2}$ was derived. A similar amount of CO$_{2}$ was also suggested at the time, since CO$_{2}$ was known to be highly abosorbing shortward of 1600 \AA. Because the energy bandwidth and total estimated \textit{dereddened} flux absorbed in this region were comparable to those in the ERE, it was speculated that the former might even be responsible for  powering the latter.

Observations of HD 44179 with greater sensitivity and spectral resolution using the \textit{Hubble Space Telescope} have have provided more detailed information about the UV flux in this object. \citet{glinski96} identified the spin-forbidden  a$^{3}\Pi$-X$^{1}\Sigma^{+}$ Cameron emission bands between 1900 and 2200 \AA, the first such detection in an astronomical object outside the solar system. The excitation mechanism for the Cameron bands remains unknown. \citet{glinski96}  suggested collisions by charged particles with energies of 7-15 eV. \citet{yan00} favored dissociative recombination of electrons with HCO$^{+}$, leaving CO in an excited a$^{3}\Pi$ state. 

\citet{glinski97}  found that fitting the 4P fluorescent (0,1), (0,2), and (0,3) emission bands required the presence of at least 2 components (and a best fit with 4), one being cold with a rotation temperature of only $\sim$50 K, and another much hotter one at $\sim$3000 K. The spectropolarimetry studied by  \citet{reese96} indicated that these emission bands also dilute the polarization, similar to what \citet{schmidt80} found for the ERE at visible wavelengths. Reese \& Sitko also drew attention to the possible presence of OH emission bands near 3080 \AA. 

\citet{glinski96} and \citet{reese96} had identified other features between 2500 \AA \ and 2800 \AA \ as possibly arising from Fe II, but offered no model to describe the emission. The presence of eleven Fe II emission lines arising from three separate multiplets at visible wavelengths \citep{hobbs04} supports the reality of these identifications. 

Because of its importance to understanding the chemical evolution of the nebula and to better understand the detailed physics of the ultraviolet spectrum, we recently undertook an effort to re-analyze the UV spectrum of HD 44179. We discuss our results in the following sections.
 
\section{Observations}

The high resolution observations of the (0-1), (0,2), and (0,3) A-X emission bands of CO were obtained on 1993 March 20 (UT) using the G160M grating and the 2.0 arcsec square Large Science Aperture of the Goddard High Resolution Spectrograph (GHRS) of the \textit{Hubble Space Telescope (HST)}. Details of the observations are described in \citet{glinski97}. 
 
All low resolution spectra were obtained using the G130H, G190H, and G270H gratings of the Faint Object Spectrograph (FOS) of the HST, with the 1.0 arcsec circular aperture centered on the star. The G190H and G270H observations were obtained on 1992 Nov. 11 (UT) in the spectropolarimetric mode, and total intensities reconstructed as described by \citet{reese96} and \citet{glinski96}. Under ideal conditions, with calibration lamp exposures obtained immediately before or after the science exposure, and without grating movement, the wavelength calibration of the FOS was generally good to a fraction of a diode. At the longer G270H wavelengths,  with a spectral resolution of 2 \AA \ diode$^{-1}$ the calibration would normally be good to $\sim$1 \AA. However, the insertion of the polarization optics introduces additional uncertainty. Usually the spectra in the two orthogonal planes of polarization had slightly different wavelength calibrations, and were simply shifted before merging (R.A. Allen, private communication), degrading the absolute wavelength calibration further. We consider the absolute calibration to be no better than 2 \AA, but the relative shift within a limited spectral region with the same grating to be more precise. These limitations will determine whether or not any conclusions can be drawn regarding the relative velocities of different emitting species. 

The G130H observations were a series of seven identical observations on 1992 Sept. 20 (UT) without the insertion of the polarimeter optics, and averaged with equal weighting. However, the archived G130H spectra were severely mangled during processing and archiving, and appear to suffer from a poor background correction. In particular, the flux levels shortward of 1550 \AA \ are inconsistent with previous spectra obtained using the \textit{International Ultraviolet Explorer (IUE)} \citep{sitko83}, as well as LiF2A spectra at $\lambda \sim 1080-1180$ \AA \ in the data archive of the  \textit{Far Ultraviolet Spectroscopic Explorer (FUSE)}, which place an upper limit on the flux at 1180 \AA \ of $F_{\lambda} \sim$ 5x10$^{-15}$ erg cm$^{-2}$s$^{-1}$\AA$^{-1}$. In Figure 1 we compare the``preview'' and final archived G130H spectra with that obtained using the \textit{IUE}. Also shown are GHRS observations from the archive obtained a few years later with the G140L grating. When smoothed to the same approximate spectral resolution as the FOS data, the overall flux levels and general spectral shapes are very similar, but the core line depths deeper.  In order to use a uniform data set throughout the entire 1300-3000 \AA \ region for much of the modeling, and to keep the differences in time of observation small, we use the FOS spectra for the entire  1300-3300 \AA \  region, except for the set of three CO bands covered by the  GHRS spectra from 1993.

\section{Model Description}

\subsection{Radiative Transfer Model}

The approach used for computing the UV transmission and emission from a gas in non-(Local Thermodynamic Equilibrium) (NLTE) is an extension of a previously developed radiative-transfer (RT) model \citep{sun93} for NLTE IR emission from high-altitude re-entry wake flows.  It employs a single-line Voigt equivalent width approximation with Curtis-Godson averaging to treat multi-segment inhomogeneous paths.  The path transmittance and radiance is computed separately for each line and a statistical line overlap correction is applied.  The computational spectral resolution is limited by the largest single line equivalent width for the spectrum under consideration.  For this study, we typically performed calculations at $\sim$1 cm$^{-1}$ resolution corresponding to $R= \lambda / \Delta\lambda > $3x10$^{4}$.  This is higher resolution than the observations, and the model predictions were degraded to lower resolution, accounting for both velocity dispersion and sensor spectral resolution.  The accuracy of this approximate RT approach is generally within ±10\% of the more accurate, but several orders of magnitude slower, monochromatic line-by-line method.  Given the simplifying assumptions we used in the spectral analysis of the highly spatially complex RR, this level of accuracy for the RT model is appropriate.

The primary extensions of the IR model for UV applications are the generalizations of the NLTE source function and line strength.  UV transitions involve upper and lower electronic states each with its own vibration-rotation energy level manifold.  The total energies of each upper and lower transition level are factored into their separate electronic, vibrational, and rotational contributions.  Populations are required for each of these six components.  It is convenient to characterize the population for each degree of freedom in terms of an effective temperature.  This often works well, however, an arbitrary vibrational level population distribution can be specified if required.  In addition to the six effective temperatures for the internal degrees of freedom, a translational temperature is also specified, which governs the Doppler line width.  In this study, the effect of velocity dispersion was introduced in terms of an effective translational temperature.

\subsection{Line Parameters}

Energy levels for the CO A$^{1}\Pi$  state were determined from a de-perturbed expression \citep{kur76}.  Although many of the A$^{1}\Pi$ state energy levels are perturbed by nearby triplet states, these perturbations are not important at the spectral resolution of this work.  Dunham coefficients were used for the energy levels of the X$^{1}\Sigma$ ground state \citep{gu83}.  The line strengths were based on Franck-Condon factors, r-centroids, and H\"{o}nl-London factors taken from Kurucz's compilation.  The relative electronic transition moments were obtained from DeLeon's quadratic representation of the r-centroid \citep{del88}.  The absolute transition moments were tied to measurements of the radiative lifetimes \citep{fie83}.  Transitions for v$^{\prime}\leq$ 23, v$^{\prime\prime}\leq$ 37, and j$\leq$130 were included in the line strength data base. 

The CO Cameron line parameters used in the spectral fitting are based on those previously used in the analysis of laboratory flame \citep{burke96} and Space Shuttle plume \citep{dimpfl05} observations of CO Cameron emission.  The line parameters correspond to vibrational transitions between the 0$\le$v$^{\prime}$$\le$7 and 0$\le$v$^{\prime\prime}$$\le$9 levels and for rotational levels up to j$^{\prime}$,j$^{\prime\prime}$$\le$100.  The Fe II line parameters were taken from the NIST atomic line compilation.\footnote{NIST Chemical Kinetics Database -  http://kinetics.nist.gov/kinetics/}  A simplified emission model was adopted for the Fe II lines, principally as a means of demonstrating the close correspondence of data features to the known Fe II lines.  The Fe II emission was modeled as optically thin and characterized by an effective emission temperature.  While it is understood that neither of these assumptions holds true for the Red Rectangle, they suffice to unambiguously confirm the presence of strong Fe II emission. 

The OH energy levels, transition frequencies, and H\"{o}nl-London factors were determined based on the Dieke and Crosswhite formulas \citep{die62}.  Corrections to the line strengths due to vibration-rotation interactions were based on the analysis by \citet{ank67}.  Frank-Condon factors calculated by \citet{alb89} were used and verified against previous calculations by others \citep{gol81}.  Absolute normalization of the line strengths were based on a well-established experimental value for the (0,0) oscillator strength \citep{rou73}. All bands with  v$^{\prime}\leq$ 4 and v$^{\prime\prime}\leq$ 2 were included, and the maximum value of the rotational quantum number for each v$^{\prime}$ level was constrained to be below the OH(A) state dissociation limit of  2 eV.

\subsection{Spectral Fits}

The general approach adopted for spectral fitting of the GHRS and FOS data is briefly reviewed, with more details presented below for the specific data features.  In all cases, a spectral baseline contribution, as indicated in Figures 2 and 3, was first subtracted.  The binary stars are obscured from direct view by a physically thick and optically opaque dust disk, and the baseline contribution corresponds to scattered star light that has not passed through the strong molecular absorption regions associated with the conical outflows.  The baseline subtracted FOS flux in the 1300-1600 \AA \ region, which is dominated by CO(A-X) absorption, was modeled as  the product of an effective source function and a computed transmittance, taken as the product of single, homogeneous segment optically thick hot and cold transmittance components.  The primary parameters determined for this spectral region are the source function, the column densities of CO, and the effective vibration and rotation temperatures of the hot and cold CO(X) ground state components.  The baseline subtracted FOS and GHRS data in the 1600-2200  \AA \ region, which is dominated by CO(A-X) emission, was modeled as a linear combination of optically thin, single, homogenous segment hot and cold emission components.  However, significant optical opacity is required to account for the attenuation in one of the cold emission components.  This emission arises from the re-emission to longer wavelengths of the absorbed photons below 1600  \AA \ (fluorescence).  The primary parameters determined for the CO emission region include the effective vibration and rotation temperatures for the CO(A) state of the hot and cold components, and a very substantial wavelength-dependent extinction attributed to dust associated with the RR.  Because the extinction curve cannot be determined directly (due to the heavy blanketing of molecular features), its properties must be estimated in order to derive self-consistent results for the lines. Additionally, the higher resolution of the GHRS data enables the determination of velocity shifts for the hot and cold components.  The baseline subtracted FOS data in the 2900-3300  \AA region, which is dominated by OH(A-X) emission, was modeled as a linear combination of optically thin, single, homogeneous segment hot and cold emission components.  The primary parameters retrieved for the OH emission include  the effective vibration and rotation temperatures for the OH(A) state of the hot and cold components. As found for CO, optical opacity is also required to account for the attenuation of one of the cold emission features. \\

\subsection{GHRS CO(A-X) Fluorescence Emission Band Fits}

The fits to the GHRS data are displayed in Figure 4.    For comparison to model calculations, it was convenient to smooth the spectrally over sampled data to somewhat lower spectral resolution.  The effective resolution of the smoothed data was $\sim$0.18 \AA \ ($\sim$7 cm$^{-1}$).   The extracted vibration and rotation temperatures for the CO(A) state of the single hot component, responsible for the wide emission base,  and their estimated uncertainties are T$_{v^{\prime}}$=20,000$\pm$5,000 K and T$_{r^{\prime}}$=5,500$\pm$500 K.  The retrieved temperatures for the single cold component, which produces the narrow emission core,  are T$_{r^{\prime}}$=50$\pm$10 K and T$_{v^{\prime}}$=2,600$\pm$200 K. The vibrational temperature was determined from the intensity ratio of the (1,3) and (1,4) features.  

It is noted that the cold component is almost completely missing from the (0,1) data, and is attributed to self-absorption.  This requires that the cold CO(X) have a sufficiently high vibration temperature to populate the v$^{\prime\prime}$=1 state in order to produce significant absorption opacity, while, at the same time, leaving the opacity of the higher v$^{\prime\prime}$ levels optically thin.  The model fits in Figure 4 show the limiting extremes of no attenuation and complete attenuation of the (0,1) cold feature. The data displays a small residual (0,1) cold feature, corresponding to $\sim$90\% attenuation of the optically thin limit.  We  estimate that a T$_{v^{\prime\prime}}$ in the 500-1,000 K range and a column density of cold CO of order $\sim$10$^{16}$cm$^{-2}$ is consistent with the data. 

The modeled  hot component is blue shifted by 294$\pm$20 km/s (-1.6$\pm$0.1  \AA)  at 1600  \AA, while no shift was required for the cold components.  In order to fit the data, we had to further reduce the effective spectral resolution of the model calculations by applying a triangular slit function of width 0.32 \AA \ ($\sim$12.5 cm$^{-1}$) attributed to a velocity dispersion of 28 km s$^{-1}$. This is somewhat higher than the upper limit of 15 km s$^{-1}$ observed in molecular lines at visible wavelengths \citep{hobbs04} suggesting we are sampling material closer to the central illuminating source. 

As an example of the spectral fitting process, the individual spectral components associated with the (0,3) fit are depicted in Figure 5.  We used the same baselines as that estimated from the FOS data as shown in Figure 3. Because of the very high effective vibration temperature for the hot component, there are overlapping contributions from a number of (v$^{\prime}$,v$^{\prime\prime}$) bands, not just the (0,3).  As a result, the hot component displays a complex and somewhat irregular line structure.  While the cold component gives rise to the most distinct feature, the integrated intensity is dominated by the hot component.  The general shape of the hot component is principally determined by the CO(A) rotational temperature, while the underlying continuum (i.e., not the baseline) is largely determined by the CO(A) vibrational temperature. It is also seen that the relatively sharp band edges for the hot bands allows for a good determination of the observed blue shift.  This implies that if emission of comparable intensity from a corresponding red shifted component were present, it would be clearly discernible in the data.  A possible origin of the one-sided velocity shift seen only for the hot emission component is discussed later.  

Inspection of Figure 4 shows that the integrated intensity of the (0,3) spectral region is approximately a factor of 1.5 times higher than that for the (0,1) region.  The relative integrated intensities of the CO(A-X) emission in these spectral bands are primarily controlled by the (0,1) and (0,3) Franck-Condon factors, since both regions arise from the same upper electronic state vibrational level.  This would predict an intensity ratio of (0,3) to (0,1) of 0.7 as opposed to the observed 1.5, a factor of 2 discrepancy!  We attribute this strongly wavelength-dependent effect to several potential causes, including, strong dust extinction in the ``hidden'' region in the general vicinity of the binaries, and the spectral albedo of the scatterers in the conical outflows  This can be expressed as a reddening function that is applied to the CO(A-X) spectral predictions,

\begin{equation} 
f_{red}=exp(0.006(\lambda-1600))
\end{equation}

\noindent where, for convenience in the spectral fitting, we normalized the function to unity at 1600  \AA \  (this technique is analogous to using the Balmer decrement in ionized gas clouds to derive the extinction characteristics of the dust there.)  Physically, this means that the dust opacity increases by a factor of 3.6 in going from 2200  \AA \ to 1600  \AA, corresponding to the full spectral range of the CO(A-X) emission.  This translates into a factor of 36 change in intensity.  This large effect will be borne out subsequently when the FOS CO(A-X) emission fit is presented, which encompasses the full range of the emission spectrum.

\subsection{ FOS CO(A-X) Fits 1300-2400  \AA}
	
The fits to the FOS in the CO(A-X) absorption, 1300-1600  \AA, and emission, 1600-2400  \AA,  spectral regions are depicted in Figure 6.  The FOS emission region was modeled with the same parameters and spectral components as determined from fitting the GHRS data. The GHRS derived reddening function was also used for the FOS fit.  The dramatic influence of the reddening on the entire emission spectrum is evident in Figure 7.  Without reddening, the long wavelength tail of the CO emission spectrum would not be easily discerned beyond $\sim$2000  \AA.

For fitting the absorption region, 1300-1600  \AA, we also used a two component, hot and cold, model. The velocity shifts were constrained by those found from the GHRS emission analysis.  However, it is noted that the effect of the velocity shift on absorption is opposite to that for emission; that is, the absorption is red shifted when the emission is blue shifted. The spectral fit in the absorption region was divided into two spectral pieces because of the strong discontinuity of the apparent source and baseline contributions at $\sim$1450  \AA \ (see Figure 1).  This sudden drop in flux is consistent with spectral models of the AGB component of the Red Rectangle \citep{men02}.   The CO(X) ground electronic state temperatures for the hot component were found to be T$_{v^{\prime\prime}}$=T$_{r^{\prime\prime}}$=2,000$\pm$500 K.  The retrieved CO column density was 3$\pm$1x10$^{17}$ cm$^{-2}$. The cold component parameter values are not well determined from this low resolution data, as the cold component only accounts for a small sliver of the spectrum near the strong band edges.  However, it is required in order to fill in missing intensity near the band edge due to red shifting of the hot component.  Based on the GHRS fit for the cold component, we used T$_{v^{\prime\prime}}$=1000K and T$_{r^{\prime\prime}}$=50K.  A column density comparable to that for the hot component was required in order to account for the missing level of absorption (see middle plot of Figure 6.).   The $^{13}$CO$/^{12}$CO isotope ratio was estimated to be 0.05, but this is not a well determined parameter; it was slightly preferred over 0.02 and 0.10.  A velocity dispersion corresponding to a Doppler temperature of T$_{D}$=2x10$^{5}$ K ($\sim$18 km/s dispersion) was required to obtain a sufficient level of absorption for the highly saturated CO features.  The actual velocity dispersion may well be higher, but at this value the individual CO lines are highly overlapped, and further line broadening has little impact on the RT calculations.  For simplicity, a constant source function was assumed and adjusted for a best fit.  The source function is certainly more complex, as it is  comprised of both the true stellar source function as well as re-emitted light from the initially absorbed photons (see Figure 6 for an estimate of the re-emission spectrum in the absorption spectral region).  Most of the absorption in the 1440-1600  \AA \ is accounted for by the CO(A-X) bands, and the remaining absorption correlates well to other known bands of CO, in this case, the spin-forbidden a$^{\prime}$-X, d-X, and e-X bands, which are not included in the model, but whose wavelengths are indicated in Figure 6.  Their presence at significant strength is confirmed by laboratory measurements of CO absorption \citep{ros00} which are compared to the FOS data in Figure 8.  The laboratory measurements were performed at a much higher pressure, 50 torr, lower temperature, 298 K, and much higher column density, 2.4x10$^{19}$ cm$^{-2}$, than exist in the Red Rectangle.  As a consequence, they correspond to a higher degree of opacity and saturation.

The spectral fit in the 1300-1450  \AA \ region, using the same parameters as for the 1440-1600  \AA, except for a lower baseline, is seen in Figure 6 to account for most of the observed absorption down to $\sim$1380  \AA \ where it begins to significantly underestimate the level of absorption.  However, while incomplete in this spectral region, the Rostas data strongly suggests that the most, if not all, of the absorption is likely attributable to CO.  

\subsection{FOS CO(a-X) \& Fe II Fits 1800-2800  \AA}

The CO(a) and Fe II model fits to the FOS data for the CO Cameron emission region are shown in Fig. 9.  For the CO(a) fit, two emission mechanisms were considered, direct photo-excitation of CO by the strong UV flux in the vicinity of the binary system, and chemical excitation which likely occurs in a different spatial region.  As discussed below, only the chemical mechanisms was found to provide a reasonable fit and is the only one displayed in Fig. 9.   For the chemical mechanism, the retrieved rotational temperature for the CO(a) emission was T$_{r^{\prime}}$=1,000$\pm$200 K.  A well defined vibrational temperature could not be determined.  Rather, an approximately equal population for all v$^{\prime}$v$^{\prime\prime}$ levels yielded the best fit.  The fit shown in Fig. 9 corresponds to equal populations for all levels.  While some improvement to the fit could be obtained by allowing for individual adjustments to the populations, this is not warranted given the as yet many uncharacterized absorption and emission features due to other species in this spectral region.  For instance, we note that the strong model-predicted (2,0) emission is not evident in the data, presumably due to the overlapping strong C I absorption feature.  For the photo-excitation mechanism, equal vibrational populations were also assumed and a somewhat higher rotational temperature of  T$_{r^{\prime}}$=3,000 K was used.  The reddening effect found for the CO(A) emission was included for the photo-excitation mechanism but not for the chemical mechanism, for reasons discussed below.  There is a good correlation of spectral features and relative intensities for the chemical mechanism.  The photo-excitation mechanism may account for some of the small residual intensity above $\sim$2550 \AA \ not accounted by the CO(a) chemical and Fe II emission channels.  However this is quite speculative and requires more careful modeling, particularly for the Fe II lines, to corroborate. 

The spectral fit to the CO(a) emission for the chemical mechanism indicates a very high degree of vibrational excitation of the v$^{\prime}$ levels.  The distribution determined here is considerably more excited than that derived previously by more approximate analysis \citep{glinski96,yan00}. The current, much hotter distribution, seems more consistent with collisional excitation by an ion, as discussed by Yan et al. However, the chemical mechanism cannot be ruled out, as significant differences in the local environment in which the chemistry occurs (i.e., Red Rectangle vs. laboratory) may effect the observed product CO(a) vibration and rotation state distributions.  The fit for the chemical mechanism did not require the reddening effect that was required for fitting the CO(A) emission.  This suggests that the chemical excitation of CO(a) occurs in a spatial region that is further away from the binaries than that associated with the direct photo-excitation that produces CO(A) and possibly some CO(a).

The modeled Fe II emission lines demonstrate a close correspondence with emission or absorption features seen in the data.  The modeled spectrum was based on an effective emission temperature of 12,000 K.  However, no great significance should be attributed to this temperature; it was merely used to adjust the relative emission intensities of the various Fe II line groups to fall within the range of the data.  It is noted that several of the stronger modeled Fe II emission lines, such as those near 2350 and 2400 \AA, correlate with absorption features in the data.  This indicates that these lines are optically opaque.  Photons absorbed in these lines escape the Red Rectangle via optically thinner transitions at longer wavelengths that originate from the same upper states but terminate on different and higher lying lower states.

\subsection{FOS OH(A-X) Fits 2750-3350  \AA}

The fit to the FOS OH(A-X) emission spectrum is displayed in Figure 9 and the fit components are shown in Figure 10.  Because of the noisy appearance of the baseline subtracted data, spectral fitting was performed on spectrally-smoothed data corresponding to a spectral resolution of $\sim$10 \AA.   The spectrum was modeled as the sum of two components, a very hot component due to UV photo-dissociation of H$_{2}$O, and a cold component due to stellar-induced fluorescence of OH.  The hot spectrum is based on a model fit to UV observations of OH(A) emission from Space Shuttle plumes \citep{ber03}.    This spectrum is not well characterized in terms of effective OH(A) vibration and rotation temperatures, and corresponds to a high degree of excitation in both the vibrational and rotational energy level populations. 

The retrieved rotational temperature for the cold components was T$_{r^{\prime}}$=250$\pm$50 K.  The approximately equal contributions of the cool (0,0) and (1,1) components are consistent with stellar pumping of vibrationally cold OH(X); this is also consistent with the lack of observed (2,2) and (2,1) cold features.  Typically, the (1,0) feature, which occurs around 2850  \AA, is observed with about a third of the intensity of the (1,1) feature.  The fact that it is not seen in the RR spectrum indicates that the (1,0) transition is optically thick.  The cold v$^{\prime}$=1 state is excited by stellar pumping of the (1,0) transitions; however, the re-emitted photons preferentially ``escape'' via the optically thin (1,1) transition.

\section{Discussion of the Modeling}

\subsection{Hot CO(A) Emission Vibrational and Rotational Temperatures}

The extremely high retrieved CO(A) emission temperatures, T$_{v^{\prime}}$=20,000 K, are a direct consequence of the efficient stellar pumping to high v$^{\prime}$ levels.  From inspection of Figure 6, it is evident that all of the CO absorption bands are highly saturated.  This means that the populations of the CO(A) v$^{\prime}$ levels will be determined more by the available stellar flux as opposed to the inherent oscillator strengths.  It is clear that levels beyond v$^{\prime}$=9, the highest observed level, are also significantly populated.  In terms of an equivalent temperature the v$^{\prime}$=9 level corresponds to 17,000 K.  Thus the retrieved effective vibrational temperatures are in line with expectations based on the stellar pumping.  Upon reflection, we should have adopted a two-temperature model for the vibrational distribution because of the strong discontinuity of the source flux at 1450  \AA, although it is noted that the single temperature model yielded a respectable fit to both the FOS and GHRS data.

The CO(A) state rotational temperature retrieved from the hot component of the emission data was T$_{r^{\prime}}$=5,500 K, in contrast to the much lower temperature retrieved for the CO(X) absorption data (the CO(A) excitation source) of  T$_{r^{\prime\prime}}$=2,000 K $-$ a discrepancy of 3,500 K!  The resolution of this apparent discrepancy is also attributed to the strong saturation effects for the absorption.  Analogously to the v$^{\prime}$ population distribution, the population of the associated j$^{\prime}$ levels will also be governed more by the available flux than their inherent line strengths.  The effect of saturation is to enhance the population of high j$^{\prime}$ levels relative to lower j$^{\prime}$ levels.  This enhancement is then directly observed in the emission spectrum, because these transitions are optically thin.

\subsection{Previous GHRS Model}

\citet{glinski97} previously analyzed the GHRS CO(A-X) emission spectrum.  They employed a four component model consisting of a first principles spectroscopic model for the hot and cold rotational emission components and empirical Gaussian models to account for cold absorption and broad emission features near the band origins.  Using these components and allowing for separate velocity shifts and dispersions for each, semi-quantitative agreement with all three GHRS spectral regions was obtained. The current first-principles based model results in an improved fit based on just two, hot and cold, emission components.   Because of the semi-empirical nature of the previous approach, it is not possible to compare it in detail to all aspects of the current modeling approach.  However, it is noted that both models use the same rotational temperature for the cold emission features.  The rotational temperatures for the hot emission component both indicate a high temperature but are significantly different $-$ 3,000 K  versus 5,500 K  $-$ for the former and present model.  In part, this difference is related to the assumption in the earlier model that only a single (v$^{\prime}$,v$^{\prime\prime}$) band contributes to each of the GHRS spectra, where as the current modeling indicates that multiple bands contribute to each.  The previous approach found a significant cold absorption component for each spectrum, where as the current approach found cold absorption only for the (0,1) band.  This difference results in very different estimates of the CO(X) ground state vibrational temperatures, $\sim$5,000 K versus 500-1,000 K for the former and present model.  Very different velocity shifts were retrieved for the two models,  for the hot emission component, +0.1  \AA \ versus -1.6  \AA \ for the former versus current model.

\subsection{Geometry of the Emitting Region}
 
Figure 11 depicts a radiative-transport geometry that is consistent with the different observed velocity shifts for the absorption by the hot CO below 1600 \AA, and the hot and cold CO emissions above 1600 \AA.  The overall geometry is based on that deduced from optical scattering data by \citet{wae96} and the Hubble observations \citep{cohen04} which indicate an opening angle of 40$^{\circ}$  for the conical flows.  The binary system is obscured from view by a thick dust disk.  In the optical spectral region, all the observed light has been attributed to dust scattering in the un-obscured portions of  conical outflows  \citep{wae96}. The direct emission from the source species is not observed and presumably originates in the obscured region.  We speculate that there is a very fast CO outflow, $\sim$300 km sec$^{-1}$, that originates from and is localized near the binary system.  Assuming that the CO absorption in the portion of the conical outflows not obscured by the disk is optically thin, then the CO emission from the stellar-excited fast CO can only be observable via scattering in the outflow.  Because of the pinched bi-conical geometry, only photons emitted in a narrow cone, $\sim \pm$20$^{\circ}$, aligned with the CO flow are available for scattering into the observational line of sight.  The scattering process will preserve the velocity shift due to the fast CO, resulting in a blue-shifted hot emission component.  Any hot redshifted absorption would be produced deep in the cavity more than 10$^{\circ}$ couterclockwise to the upward normal in the figure, and would likely go undetected.  We also require the presence of a much colder and slower CO flow.  This will lead to cold emission features with small to negligible blue shifts and corresponding absorption features with small to negligible red shifts.  This is consistent with the observation that the sharp absorption band edges of the CO appear to be un-shifted, where as the emission band edges appear highly blue-shifted. 

In Figure 13 we compare the GHRS spectrum of the (0,2) A-X band with the polarization reported by \citet{reese96}. It is clear that the slow cold CO band and C I band are located at the polarization minimum, indicating that they are undergoing less scattering than the hot fast CO, whose higher polarization indicates production deeper within the scattering region. As illustrated in Figure 12, photons emitted in the direction of the observer are blocked by the dust disk. 

We note the remarkable similarity of this RT geometry to the dynamical outflow model proposed by \citet{sok05}.  Soker proposes that there are intermittent periods of enhanced  high stellar mass loss that can produce bi-polar jets localized within $\sim$1-2 AU of the binary system, with a velocity of $\sim$300 km/sec.  However, most of the conical volume in the disc-obscured region is filled with much colder and slower velocity gas, $<$10 km/sec, from periods of ÒnormalÓ mass loss.  

\section{Other Consequences}

\subsection{The CO Fourth Positive (4P) Absorption System}

The entire spectrum of HD 44179 is a complex mixture of overlapping emission and absorption features. As was  demonstrated above, these arise from a complex of highly excited electronic, vibration, and rotation levels, all of which are interconnected by allowed and (normally) ``forbidden''  a$^{\prime}$-X, d-X, and e-X transitions, which are seen  \textit{in absorption}. The presence of these features in significant strength indicates yet one more possible mechanism capable of exciting the a$^{3}\Pi$ state that gives rise to the Cameron bands: radiative cascades from the a$^{\prime 3}\Sigma^{+}$, d$^{3}\Delta_{i}$, and e$^{3}\Sigma^{-}$ states to the a$^{3}\Pi$ state (usually referred to as the CO Triplet bands). A fully self-consistent model will need to include a complete treatment of the radiative transfer that includes all of these  transitions, but that is beyond the scope of this investigation. 

One of the consequences of the tremendous optical thickness of the 4P bands near the low rotational states that dominate the band cores is that transitions back down from the excited A electronic state  to the ground X electronic state produce photons that can easily excite nearby CO molecules right back up into the A state. Like Lyman $\alpha$ photons in a gas containing neutral H, these photons will simply ``rattle around'' the gas until there is another way to remove the photon: absorption by dust or another molecule, or by branching into another less-populated electronic state. The presence of a plethora of spin-forbidden absorption features indicates that CO is probably capable of converting some fraction of the energy absorbed in the 4P system into transitions among a variety of other electronic states.

\subsection{The Maze}

In Figures 2 and 3  we show the entire UV spectrum of HD 44179. At 2200 \AA \ there is a distinct change in slope of the spectrum, coincident with the beginning of the CO emission blends, beginning with the Cameron bands. The dashed line in the figure indicates the continuum level used for the models of the fluorescence emission. The flux shortward of 2200 \AA \ is increasingly dominated by CO emission as one goes to shorter wavelengths. Were it not for the very strong absorption features below 1600 \AA, the CO emission complex would continue to much shorter wavelengths. Because of this, one needs to think of the entire region below 1600 \AA \ as one where we are not simply seeing absorption of a stellar photosphere by CO, but rather a complex blend of both emission and absorption by CO overlying a (probably hidden) stellar photosphere. That is, much or all of the emitted photons seen in this region are due to CO itself.

\section{The Big Picture}

We have demonstrated that the UV spectrum of HD 441679 is completely dominated by a complex network of absorption and emission due to CO. There are two important consequences of this. 

First, \textit{little or no stellar photosphere is apparent in the flux shortward of 2200 \AA}.  While it seems likely that exciting UV flux may be coming from an obscured (along our line of sight) hot compact companion to the luminous A-F primary star of HD 44179, or an accretion disk deply embedded in the system, at present we cannot detect it unambiguously in the UV spectrum. Unfortunately, this means that \textit{one simply cannot derive the dust extinction at these wavelengths using the standard ``pair'' method}, as was first done by \citet{ssm81} and again by \citet{vijh05}. The wavelength region between 1550 \AA \ and 2200 \AA \ are clearly affected by a vast blend of overlapping emission bands of CO. This would make any extinction curve derived to have too little extinction at these wavelengths. Shortward of 1550 \AA, the continuation of this emission is joined by even stronger absorption due mostly to the A-X bands, which will distort the extinction curve further, most likely producing too much extinction at these wavelengths.  

Second, \textit{virtually everything shortward of 2200 \AA \ is due to CO}. This means that in order to be able to unambiguously identify other molecular species in this region of the spectrum of HD 44179, it is necessary to first understand the intrinsic spectral signature laid down by CO. In principle, many interesting organic molecules may have absorption and emission features here, but without adequate removal of the effects of CO, it will not be an easy task to identify these other species. 

In order to properly characterize the entire wealth of hundreds of lines that are present in the spectrum, it will be necessary to undertake a full radiative transfer treatment of the CO-containing region. Unfortunately, we still have no firm information about the intrinsic (stellar?) photon field that presumably excites the CO, nor are all the transition rates among all of the relevant states of CO completely known. Nevertheless, there may be some important \textit{general} consequences of the tremendous excitation seen in the CO. 

One of these is the possibility that at least some of the unidentified features seen in the Red Rectangle might arise from hitherto unobserved transitions in CO. The high optical depth of the 4P system of CO has the potential to excite a variety of other emission processes. These would include the heating of dust grains, photoexcitation of other molecules, and fluorescent pumping of other transition in CO itself. For the latter two mechanisms, highly non-equilibrium population states will result. 

It has been frequently speculated that this UV-pumping might be responsible for some of the other features seen in the Red Rectangle. \citet{sitko83} had tentatively identified CO$_{2}$ as a possible absorber shortward of 1600 \AA, and noted that the width of the feature in energy (as determined from laboratory spectra) and the total energy absorbed (after removal of a few magnitudes of extinction), were comparable to the ERE feature. Unfortunately, the massive complexity of the CO spectrum has so far thwarted any attempt to positively identify the presence of  CO$_{2}$  in the ultraviolet, although is has been detected at infrared wavelengths \citep{waters98}.

\citet{glinski96} suggested that other excited emission bands (such as the Asundi a$^{\prime}$-a bands of CO, or the comet-tail A-X bands of CO$^{+}$) might be observable at visible wavelengths, although they have not been detected yet. \citet{ben99} suggested  the radiatively-pumped excited triplet states of CO itself as the possible source of the ERE in the Red Rectangle. But the ERE is now known to be present in many astronomical sources lacking the Red Rectangle's unusual CO spectrum. It has been suggested that the ERE might represent emission from silicon nanoparticles,  PAHs (both neutral and ionized), PAH dimers (charged), Hydrogenated Amorphous Carbons (HACs), and a variety of other organic and inorganic species \citep{witt98,led01,witt04,witt06,rhee07,berne08}.   Interestingly, the 3.3 $\mu$m feature in the Red rectangle is \textit{not} well-correlated with the ERE, but with the blue luiminescence (BL) \citep{witt06}.  [Note that thin films of hydrogenated amorphous silicon carbide \citep{ma98} may be capable of producing both, simply by changing the C/Si ratio. These bands arise not from SiC, but a variety of aromatic, Si-H and C-H stretching bands, and are also narrower than the SiC band - one characteristic used by \citet{vijh04} to discount SiC as a source of the BL].  

The unusual environment near HD 44179 may also be relevant to the unidentified narrow-band features seen at visible wavelengths. These bands are found throughout much of the Red Rectangle, and their structure is dependent on distance from the star itself \citep{vw02}. Many suggestions have been made for the source of these bands, such as carbynes \citep{ws81} and PAHs \citep{dh86}. Their wavelengths are also tantalizingly close to those of the Diffuse Interstellar Bands (DIBS),  and a connection between the two has been suggested many times \citep{scarrott02,vw02,sarre06}, although a definitive connection has proved elusive. The DIBS are ubiquitous; the optical bands of the Red Rectangle are not. Is CO of any relevance?

In Figure 2 we are able to trace the A-X absorption system to at least (9,0), indicating that the v$^{\prime}$=9 level of the   A$^{1}\Pi$   electronic state is being populated. The presence of the (1,3) and (1,4) fluorescence bands of the A-X system shown in Figure 4 means that the  v$^{\prime}$=1  energy state ($\sim$66,000 cm$^{-1}$) is populated sufficiently to allow downward transitions sufficiently strong to produce the fluorescent emission. The spin-forbidden absorptions observed in the a$^{\prime}$-X, d-X and e-X are also populating energy levels close to this. We also know that the lowest energy states in the  a$^{3}\Pi$ state ($\sim$48,000 cm$^{-1}$) are populated. If these states support transitions among one another, they may produce a complex blend of features at energies starting at $\sim$5600 \AA \ (18,000 cm$^{-1}$) and continuing to longer wavelengths. This may be the origin of the narrow features seen in the Red Rectangle, at least within a few arcsec of the central stars, where CO seems to dominate all other identified species. Whether it is capable of producing emission far from the stars, including the change in spectral shape with distance, is unknown and possibly problematic. 

A second consequence is that even if CO is not directly responsible for any of the features observed at longer wavelengths, it will still be necessary to provide precursor molecules for the actual emitters. This would seem to require UV photons and one or more abundant molecules. Both are evident in the spectrum of HD 44179, where we see abundant CO \textit{and} a source of UV photons. The UV photons that we actually observe are also due to CO, although a more deeply embedded source (stellar or accretion disk) is likely to be present. 

At the present time we also have no convincing explanation for the presence of the OH bands. Certainly a source of H$_{2}$O in the form of vapor or ice would be a good source for the material. Currently, the material flowing out from the central region to form the nebula is C-rich, not O-rich. Nevertheless, the torus seems to be rich in silicate dust \citep{waters98}, possibly the result of the outflow of material from one or both stars prior to the dredge-up of the He-burning produces from its (their) interiors.  

In the model of the star system by \citet{men02},  the system began as a pair of stars of roughly 2 M$_{\sun}$ each, separated by 126 R$_{\sun}$.   In their model, the originally more massive star has  passed through the red giant phase and is now a He-rich white dwarf, while the other star has just completed a similar evolutionary path (but becoming an AGB star), and is the source for much of the current mass outflow in the system. Throughout, the separation of the pair varies, increasing from 126 R$_{\sun}$ to 589 R$_{\sun}$ and finally back to 192 R$_{\sun}$. In the process, a dust envelope of mass $M_{dust}\sim0.015 M_{\sun}$ is produced. Under the assumption that the gas shed from the evolving stars is retained in the system, a net 1.2 $M_{\sun}$ remains in the envelope. 

Some post-AGB stars are known to possess circumstellar grains of water ice. HD 161796 seems to have gone through a phase that lasted approximately 900 years, where a massive outflow condensed into grains that contained about 27\% of the dust mass \citep{hoog02}. If such a system had a secondary star which could later re-heat this material, a ready supply of water vapor would be available for photodissociation by a hot source of photons. Many post-AGB systems possess O-rich dust, where in many cases the dust is stored in an unseen companion \citep{vw03}. 

\citet{sf01} have suggested that the water vapor observed in IRC +10216 could have even been produced by the vaporization of a ``Kuiper Belt'' cometary system surrounding the star. If located far enough from the system, water ice could survive the evolution of the star into the AGB phase. In their case, a mass of ice of $\sim10 M_{\earth}$ is required. The total mass in the form of solids is likely higher, but not by a large amount. While this is significantly less than the mass of dust seen in the Red Rectangle, it does suggest that at least some fraction of the O-rich material could be ``primordial'' as well as from a stellar outflow. 

It is known that multiple star systems are capable of retaining their pre-main sequence disk material. Hen3-600 is a \textit{triple} star system, with two close stars and a more distant companion, with a small circumbinary disk surrounding the tightly-bound components, and a more distant one that is exterior to the orbit of the outer component  \citep{webb01,zuck01}. We also know that some main sequence stars with masses similar to those sometimes suggested  for HD 44179 \citep{men02}, $\sim$2 M$_{\sun}$, have retained significant dust material after 10$^{7}$ years or more \citep{df04}, a significant fraction of their main sequence lifetimes. What will happen to these disks (and planetesimals, cometesimals, and planets) when one of these stars were to inevitably evolve to become a red giant or AGB star as HD 44179 has? 

For the Red Rectangle system, with main sequence luminosities of $L \sim 11 L_{\sun}$, each star would be capable of supporting a circumstellar disk with a ``snow line''  for H$_{2}$O as close as 11 AU (using a sublimation temperature in a vacuum of 150 K), and  a more distant circumbinary disk. In the outer disk, each volatile will possess its own  ``snow line''. The changing separation of the stars, as well as the growth of the luminosity to near 6000 L$_{\sun}$ virtually insures that  any objects (asteroids, comets, or planets) within the system will undergo significant changes in temperature. As the luminosities rise, the more volatile species (CO, CO$_{2}$, CH$_{4}$, NH$_{3}$) will vaporize first, with H$_{2}$O lagging significantly behind. Gravitational perturbations due to the changing separation of the wide components may stir the system further. Whether such a system can provide the OH in sufficient quantities, along with the other positively identified species seen in the system - CH, CH$^{+}$, and CN \citep{hobbs04} - is unknown. 

Higher-quality spectra of the OH bands, and a search for rovibrational transitions in the near-infrared are needed for further clarification.

\section{Where is the Hot Exciting Star?}

The high degree of excitation in the CO bands requires a source of ultraviolet flux. The more evolved star, now the ``secondary'' star whose light has not been detected at visible wavelengths, is generally assumed to be either a white dwarf or on its way to becoming one. In the model of \citet{men02} the UV tail in the spectrum is identified with this object. However, we have demonstrated that in fact this part of the spectrum is dominated by a complex blend of CO emission, not a stellar photosphere. Nevertheless, it would be important to be able to know something about this star.

In Figure 14 we show the spectrum in the vicinity of 1930 \AA. In addition to the narrow C I band structure discussed by \citet{glinski96}, HD 44179 also exhibits a \textit{broad} feature best seen in the low-resolution FOS spectra (Figure 3). Also shown in Figure 13 is the UV spectrum of the white dwarf star in the LDS 678 binary star system, taken from the \textit{IUE} archives, and whose spectrum has been discussed by \citet{weg81}. To simulate the effects of the added emission line blanketing due to  the A-X and a-X bands of CO in this region, and possible mis-match in true C I band strength, we have diluted the spectrum of LDS 678 by a factor of 4.5 and added a constant flux of 4.4x10$^{-14}$ erg cm$^{-2}$s$^{-1}$\AA$^{-1}$.

While \citet{weg81} described LDS 678 B as a cool ($T\sim10,000 K$) DC type white dwarf, more recent work by \citet{osw91} classify it as DBQA5, and also refer to it as LDS 678 A to be consistent with Luyten's original designation. The parallax of the system \citep{har80} is consistent with a distance of 10 pc. The distance to HD 44179 that would be implied  by the scaling of LDS 678A used in Figure 13 would only be only 21 pc (closer if suffering extinction)! This is, however, far closer than any other estimate of the distance to the Red Rectangle: 330 - 710 pc \citep{men02}. The spectrum of LDS 678 A also peaks near 3500 \AA \ in these flux units, also not consistent with what is seen. Likewise, we do not see the other strong C I bands exhibited by LDS 678 A at shorter wavelengths, although these might be hopelessly blended with the CO bands. Finally, if the hot component in HD 44179 were a very luminous hot star, as suggested by Men'shchikov et al., its surface gravity would be insufficient to produce the broad asymmetric C I band that due to normal pressure effects, unless the atmosphere were unusually transparent, allowing us to see deep high-pressure layers. 

Other possible explanations for the shape of this feature is that it is the result of a wind containing C I, or that it is a red-degraded band of some as yet unknown molecular species. If the former, it would require an outflow speed v$\sim$8000 km s$^{-1}$=0.02c, which seems unlikely. The alternative is a blue-degraded band structure of some molecule which just happens to coincide with the rest wavelength of the C I line. This explanation cannot be ruled out, but hardly seems satisfactory. 

Without further information, It is difficult to assign any specific origin for this feature.

\section{Conclusions}

The UV spectrum of HD 44179 provides an unusual window into the energetics of the innermost regions of the Red Rectangle nebula. There are more  identified electronic-vibration bands in this portion of the spectrum than are known anywhere else in the nebula. Modeling the features has revealed information on the vibrational, rotational, and kinetic temperatures of the gas. The high optical depth of the dominant species, CO, is bound to have a profound effect on the emission of all other molecules that coexist with it. 

The entire spectrum at wavelengths shortward of 2200 \AA \ is dominated by a complex of emission and absorption bands due to CO with high rotational and vibrational temperatures. The column density is high enough that spin-forbidden transitions are observed in absorption. The strength and complexity of the spectrum makes deriving the extinction characteristics in this region of the spectrum virtually impossible.\

We have tentatively identified emission bands due to OH between 2800-3300 \AA. Such bands might be the result of the sublimation and subsequent photodissociation of H$_{2}$O ice, yielding OH in an excited stare. Additional emission due to photo-induced fluorescence. 

Further progress will require detailed radiative transfer calculations for CO. Additional observational progress in the UV will require the use of new instruments, such as the Cosmic Origins Spectrograph (COS) planned for the HST. 

\acknowledgments
The observations were originally obtained as part of \textit{HST} grant GO 3468. L. Bernstein acknowledges support via  the Spectral Sciences, Inc. internal research program. M. Sitko also acknowledges the additional support of the Department of Physics of the University of Cincinnati, and of the Space Science Institute. The authors appreciate technical interactions with W. Klemperer (Harvard University) regarding the origin of the CO(a) emission. We would also like to thank the anonymous referee for their useful comments.

\begin{figure}
\center
\plotone{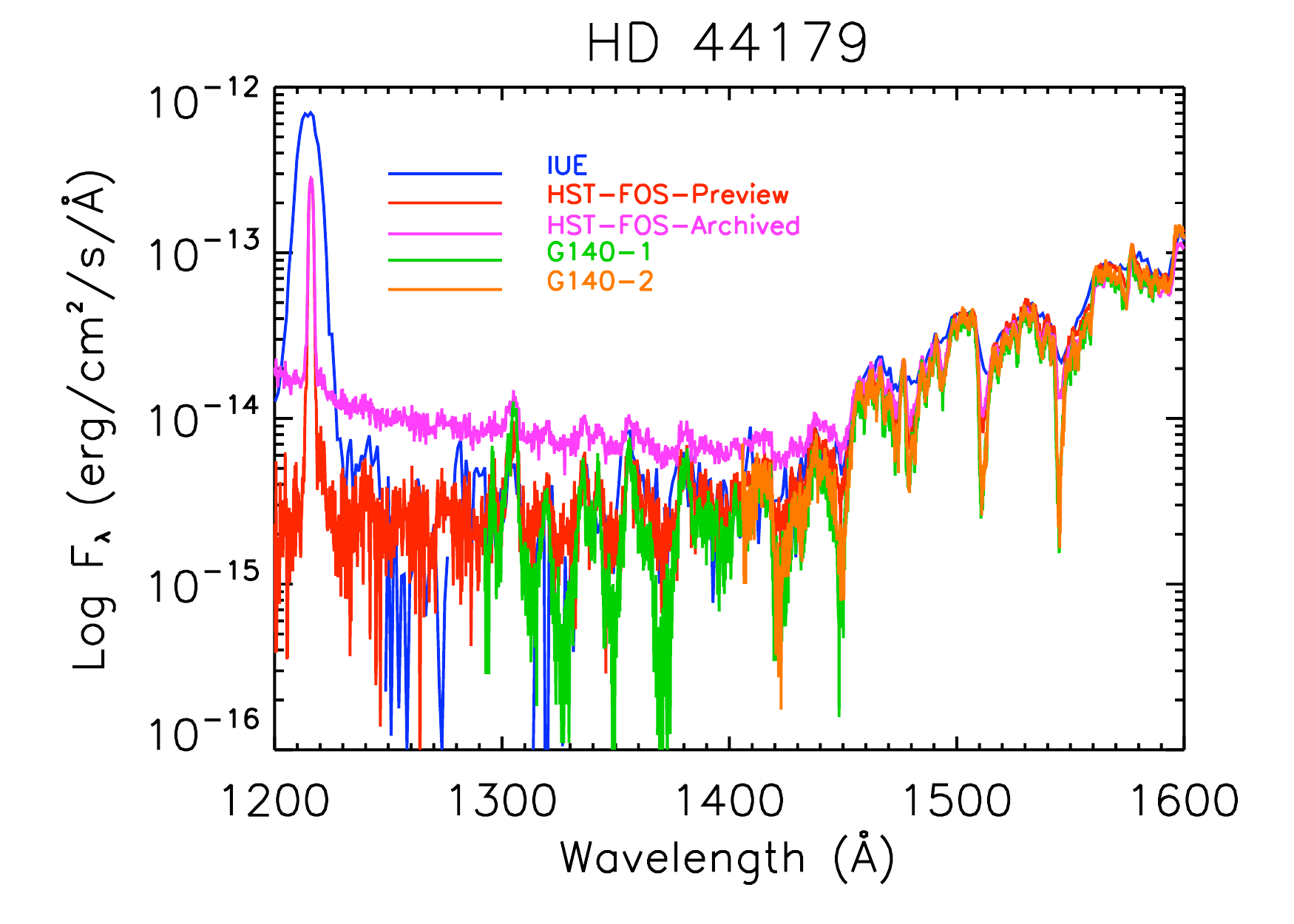}
\caption{The FOS spectrum of HD 44179 shortward of 1600 \AA. The archived spectrum included a bad flat-field correction that is missing from the ``preview''  spectrum, processed using an earlier calibration. The latter is consistent with flux levels obtained using the \textit{International Ultraviolet Explorer}, and were used for the present analysis. Also shown are two GHRS spectra obtained a few years after the rest of the data, smoothed to the approximate spectral resolution as the FOS spectra. \label{fig1}}
\end{figure}
 
\clearpage

\begin{figure}
\center
\plotone{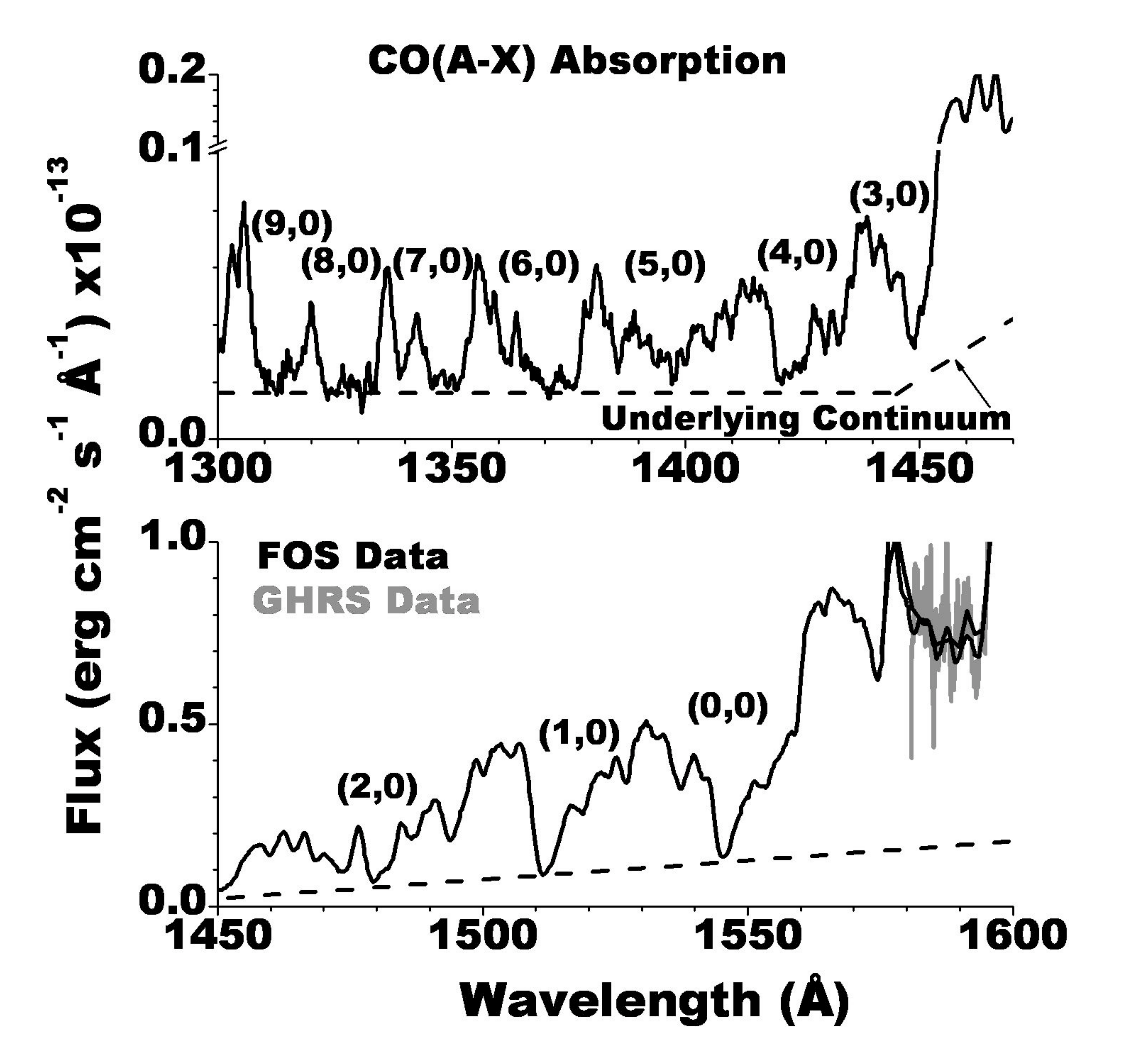}
\caption{FOS data in the CO(A-X) absorption spectral region. Also shown is the estimated underlying continuum attributed to scattered stellar light that has not traversed a spatial region of the RR containing the CO.  Above ~1560 \AA, the CO features are in emission.  A small portion of the emission region is included in the bottom panel where comparison is also made to the higher spectral resolution GHRS data.\label{fig2}}
\end{figure}
 
\clearpage

\begin{figure}
\center
\plotone{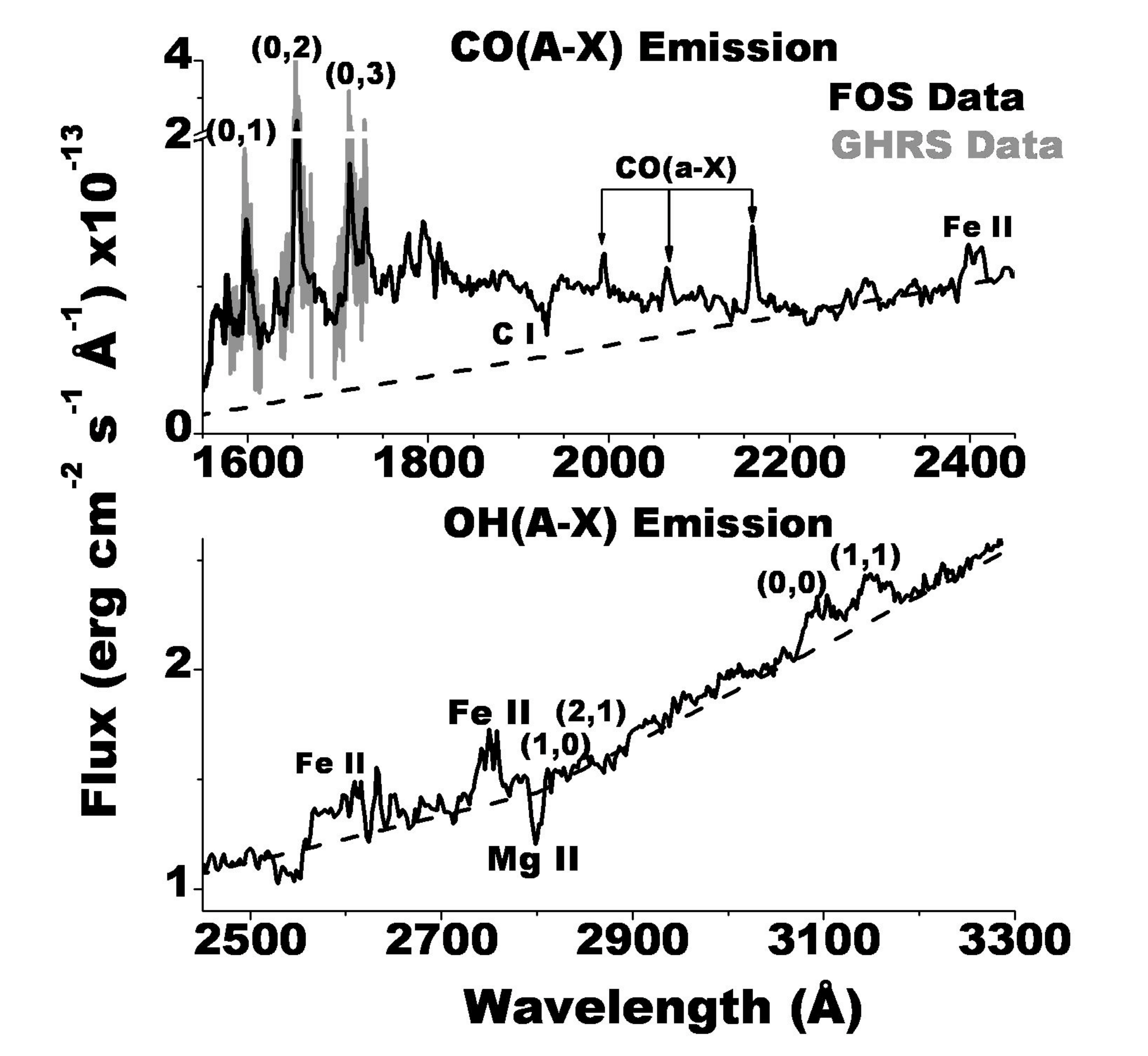}
\caption{The FOS and GHRS data are compared in the CO(A-X) emission region (top panel).  The bottom panel presents the FOS data in the OH(A-X) emission region.  Also shown is the estimated underlying continuum for the FOS data.\label{fig3}}
\end{figure}
 
\begin{figure}
\center
\plotone{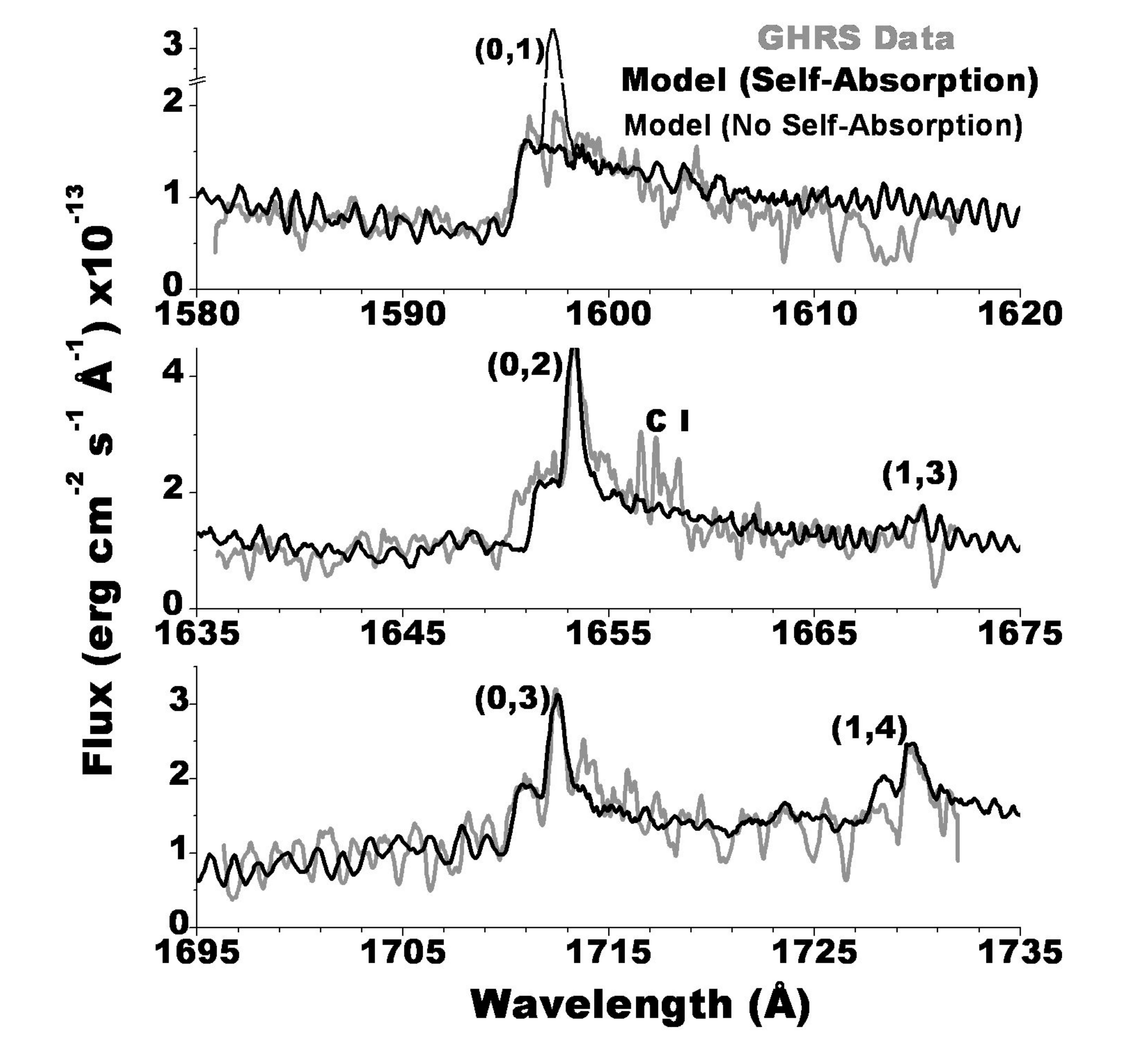}
\caption{Comparison of model fits to the GHRS data in the CO(A-X) emission region.  The upper panel shows the effect of self-absorption for the cold emission components for the (0,1) band.  Self-absorption effects are not evident for the (0,2) and (0,3) bands. \label{fig4}}
\end{figure}
 
\clearpage

\begin{figure}
\center
\plotone{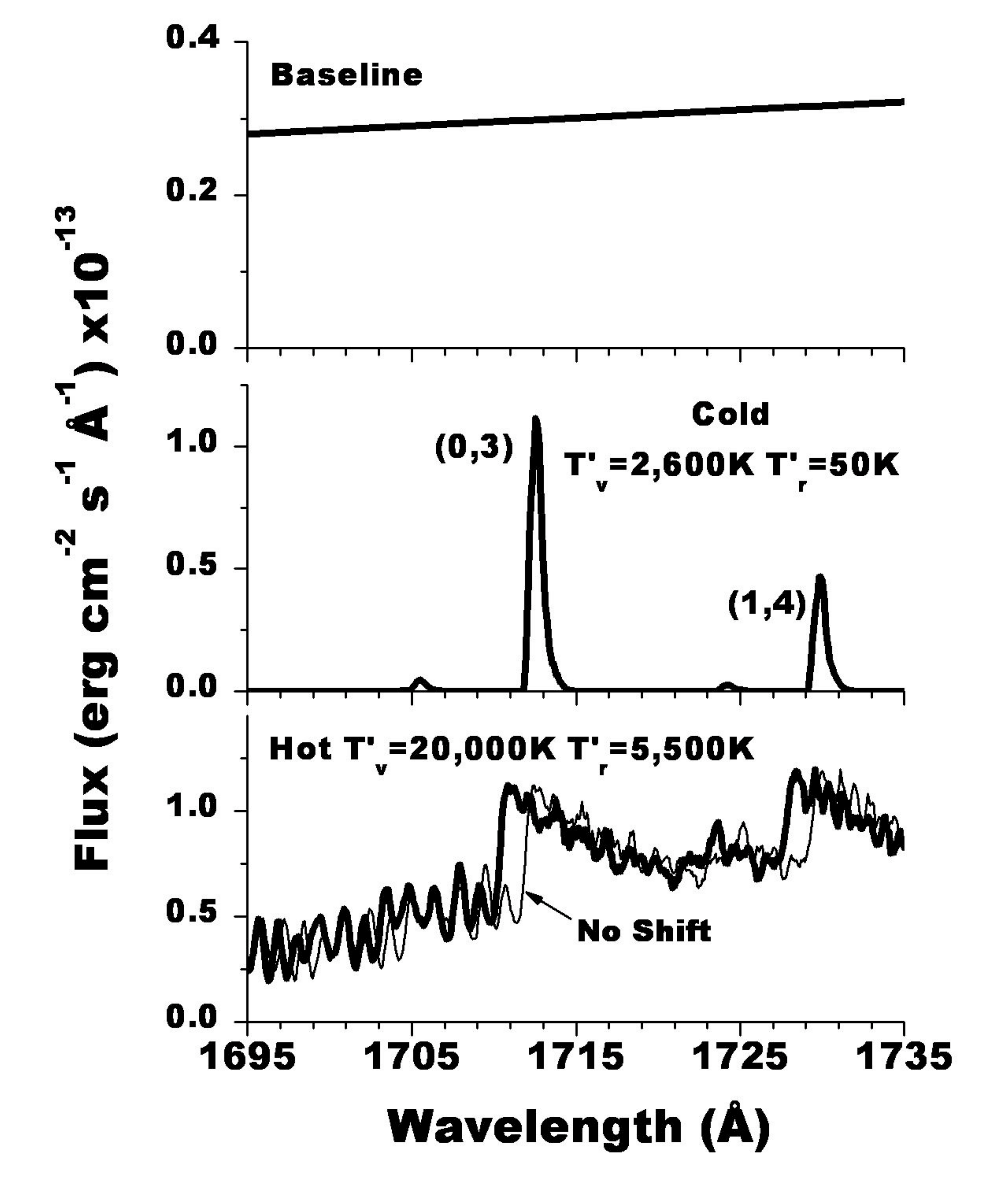}
\caption{Model fit components for the (0,3) GHRS emission spectrum.  The effect of the -1.57 \AA \  shift required to fit the data is shown in the bottom plot.  \label{fig5}}
\end{figure}
 
\clearpage

\begin{figure}
\center
\plotone{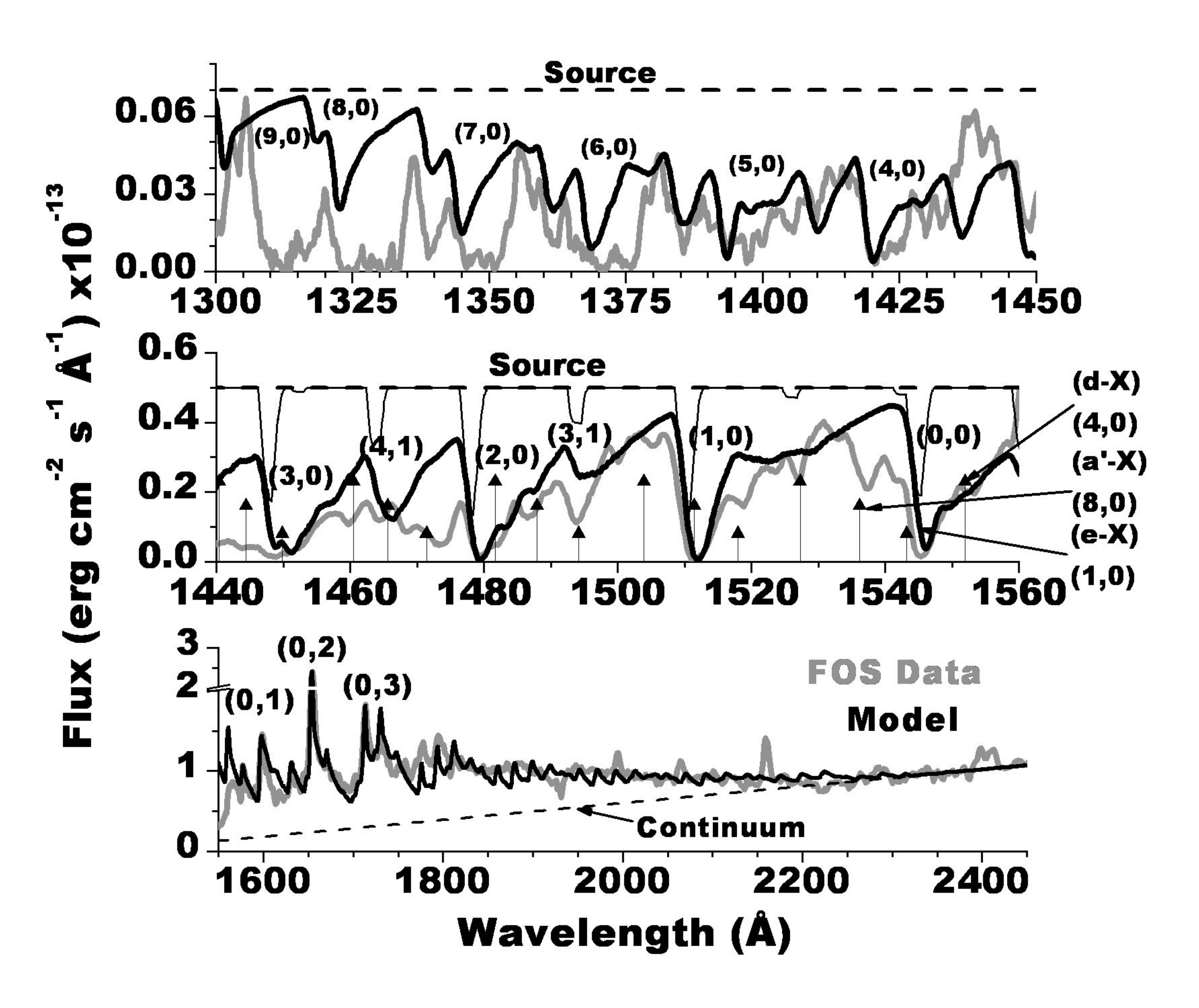}
\caption{Comparison of model fits to the FOS data in the CO(A-X) absorption (top panels) and emission (bottom panel) regions.  Baseline-subtracted FOS data was used for the absorption region comparisons.   The same constant CO column density of 2.7x10$^{17}$ cm$^{-2}$ was used for the entire absorption spectral region, 1300-1560 \AA.  The retrieved source function for each of the absorption regions is also displayed.  Only CO(A-X) absorption and emission features are included in the fits.  The (v$^{\prime}$,0) band locations of CO absorption transitions originating from other electronic states are indicated in the middle panel with a $\uparrow$ symbol.  The lowest v$^{\prime}$ band for each sequence is labeled.  \label{fig6}}
\end{figure}
 
\clearpage

\begin{figure}
\center
\plotone{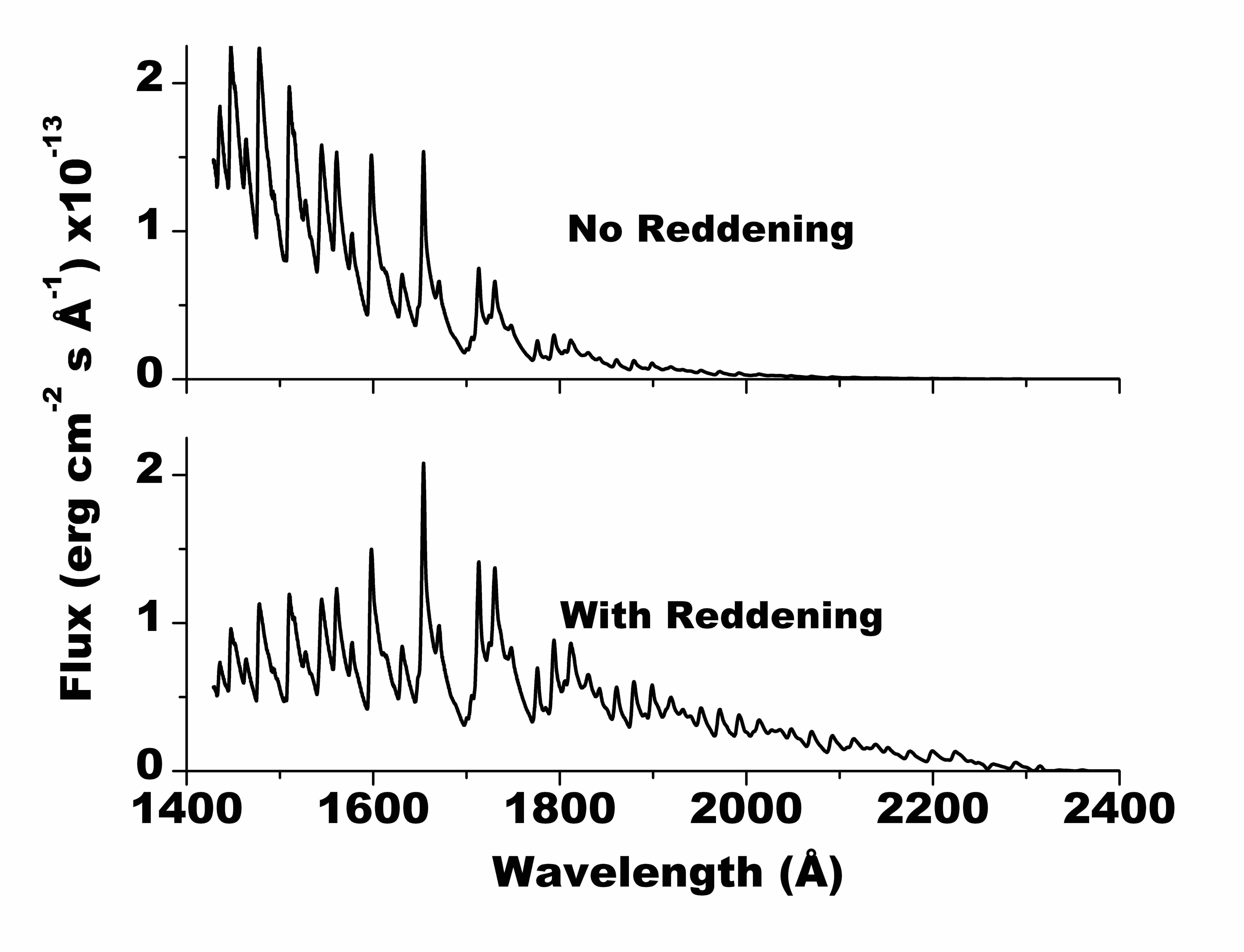}
\caption{The reddening effect due to wavelength-dependent dust extinction on the CO(A-X) emission spectrum.  Both curves are normalized to unity at 1600 \AA.  \label{fig7}}
\end{figure}
 
\clearpage

\begin{figure}
\center
\plotone{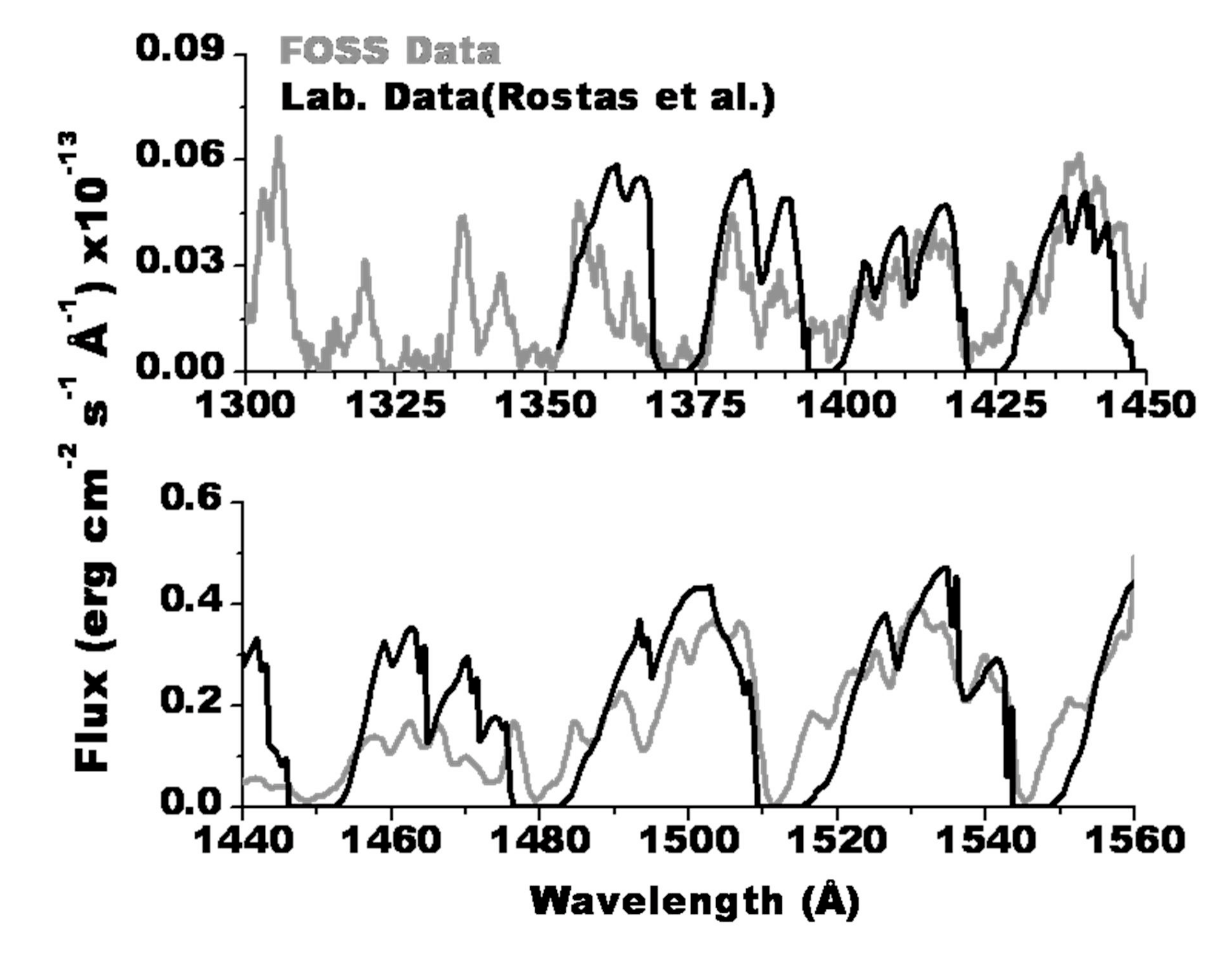}
\caption{Comparison of laboratory CO absorption measurements to the baseline-subtracted FOS data.  \label{fig8}}
\end{figure}

\clearpage

\begin{figure}
\center
\plotone{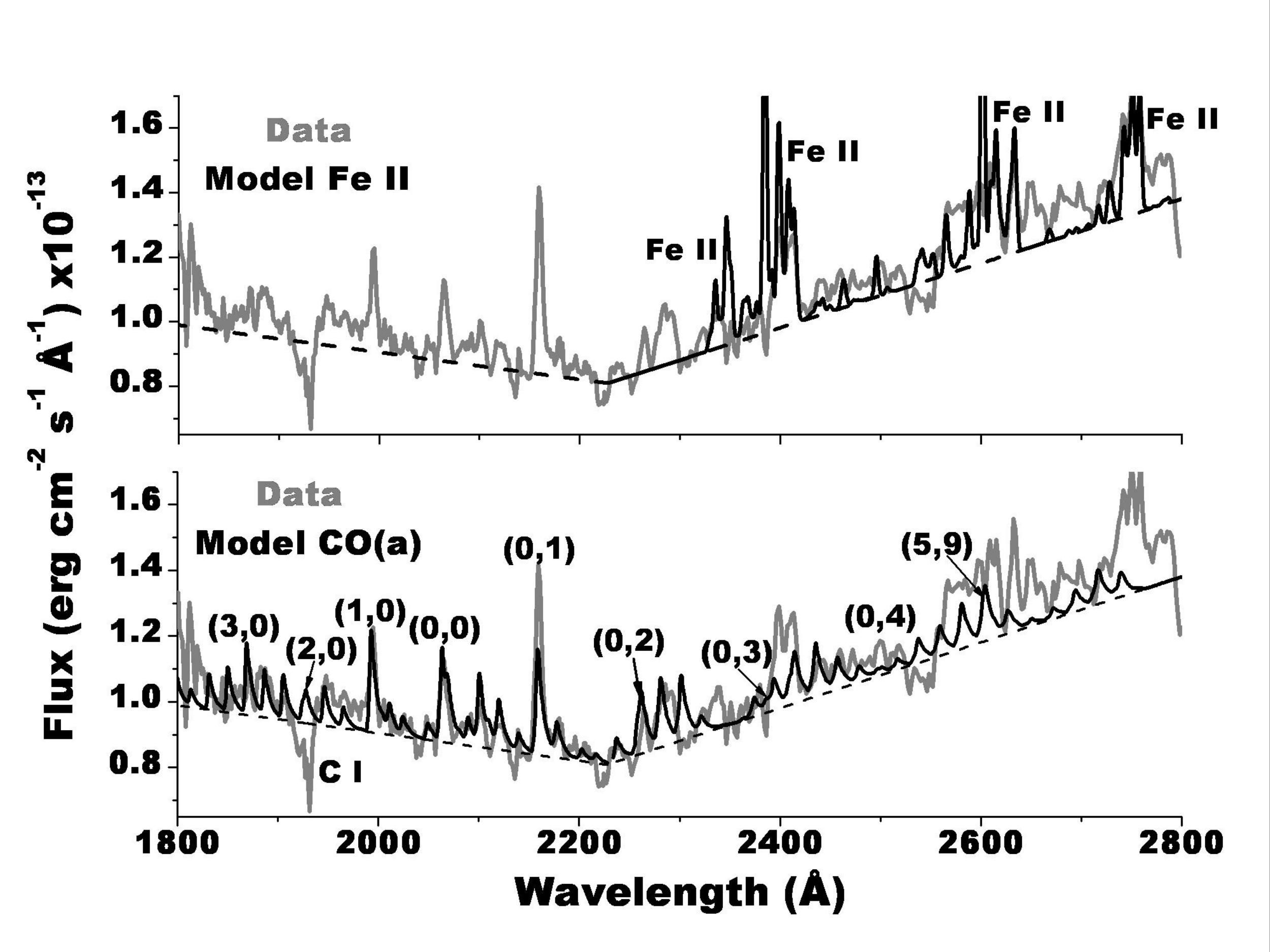}
\caption{Contributions of CO a-X and Fe II emission to the spectrum between 1800-2800 \AA. The estimated baseline used in the fitting is indicated by the dashed line. \label{fig9}}
\end{figure}
 
\clearpage

\begin{figure}
\center
\plotone{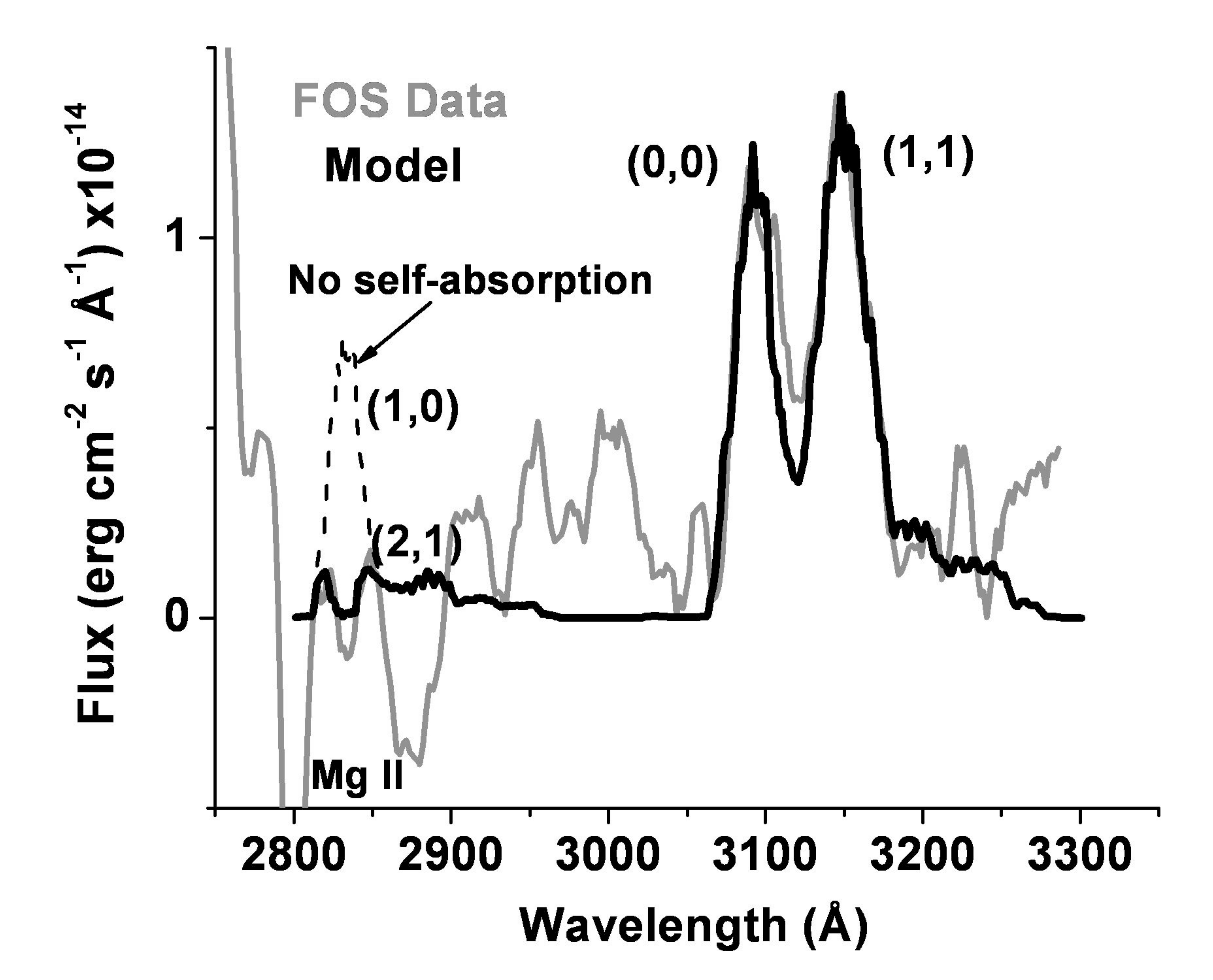}
\caption{Comparison of the model fit to the FOS baseline-subtracted data in the OH(A-X) emission region.  \label{fig10}}
\end{figure}
 
\clearpage

\begin{figure}
\center
\plotone{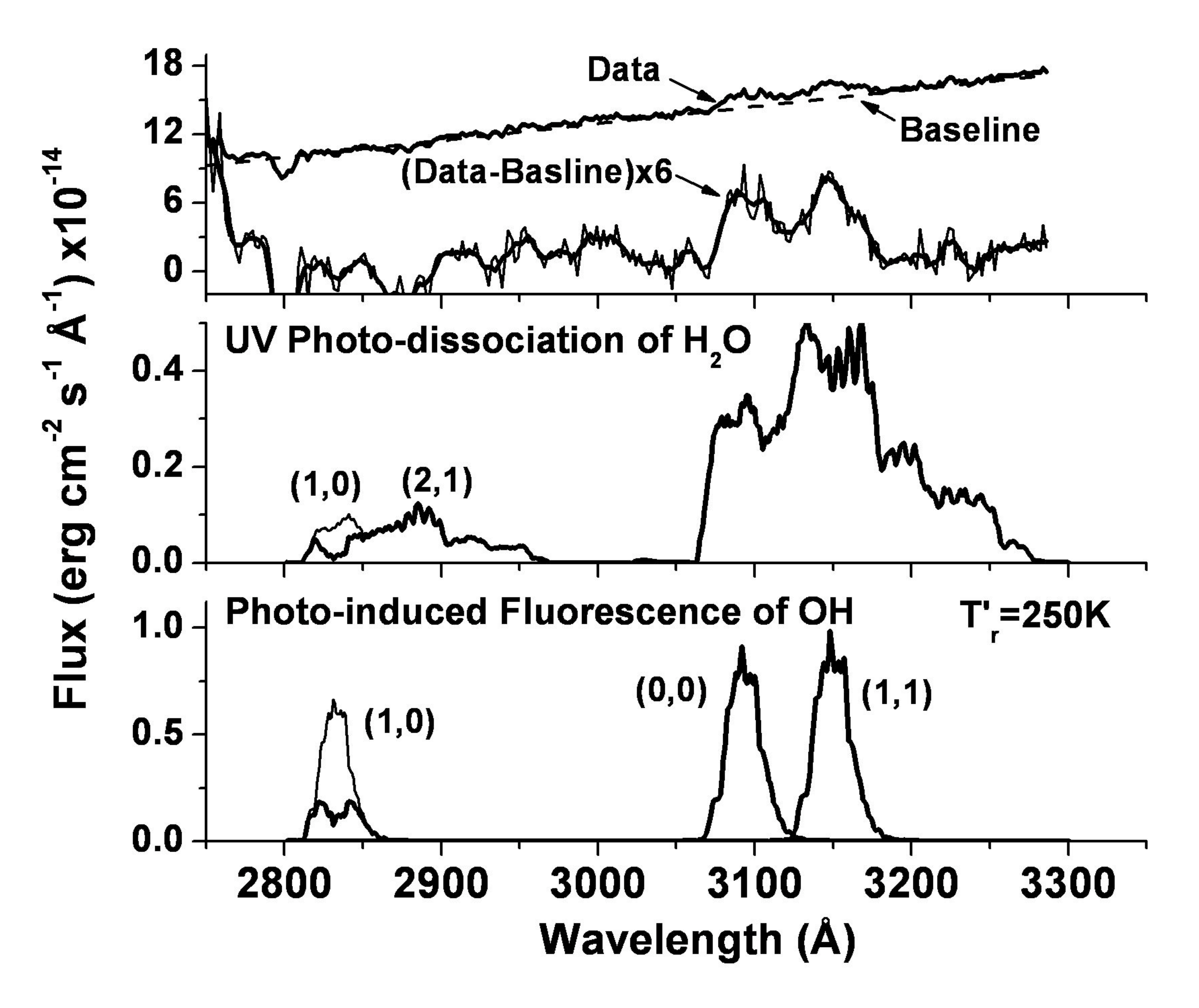}
\caption{Model fit components for the FOS OH(A-X) emission spectrum. The thin black curves for the middle and bottom panels are correspond to the limit of no self-absorption.\label{fig11}}
\end{figure}

\clearpage

\begin{figure}
\center
\plotone{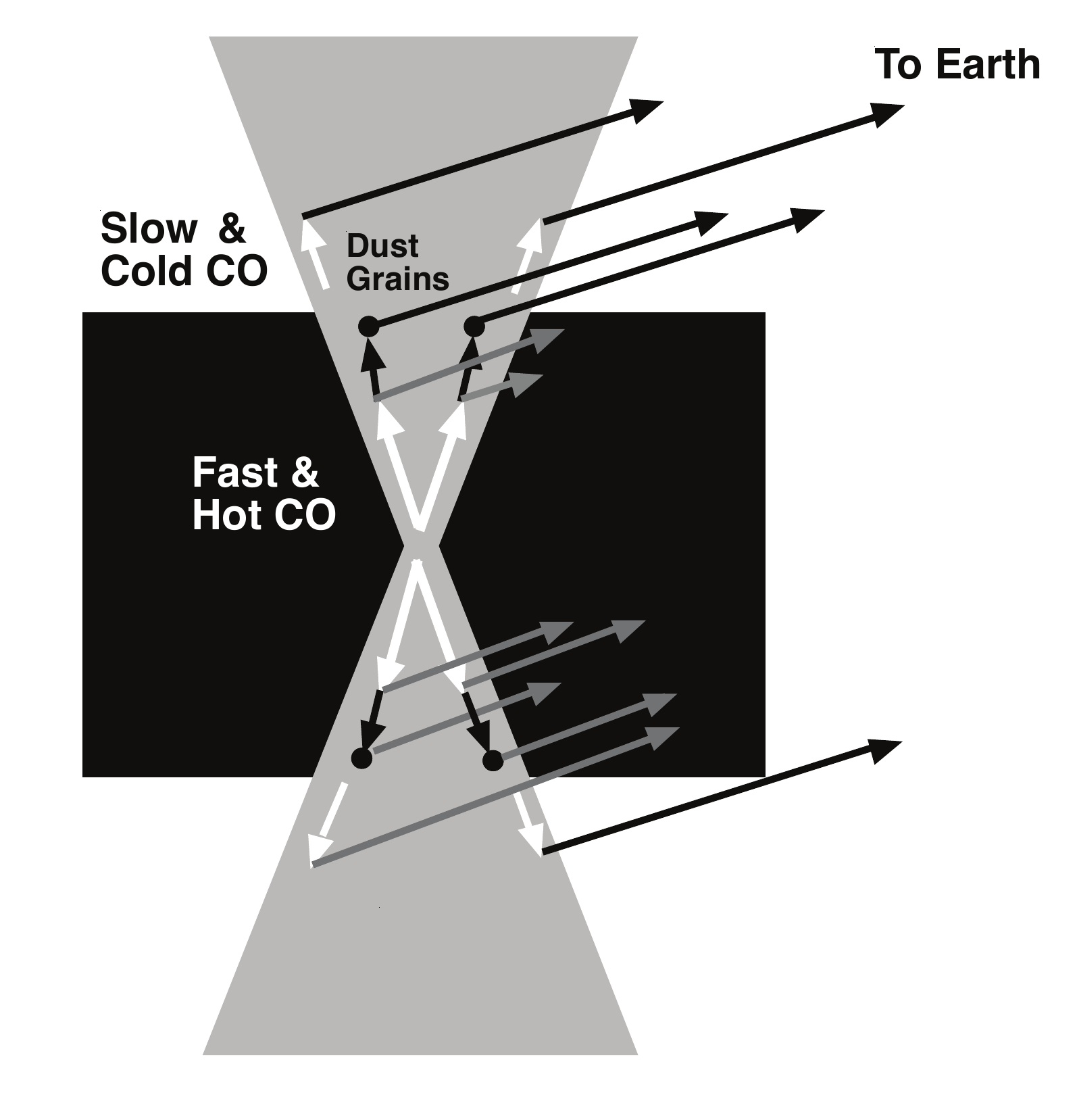}
\caption{The presence of both fast hot CO and cold slow CO may be due to deceleration in the molecular outflow. Here, the hotter higher velocity material is emitted close to the star, and is significantly occulted by the dust torus.  Much of the emission that reaches the observer only does so by way of scattering off of lower density dust mixed in with the outflowing gas. The slower-moving cooler material is produced further from the star, and is more easily seen without significant scattering. The images in the H and K bands of \citet{men02} clearly show scattered light is visible from both above and below the mid-plane of the torus. However, the extinction in the photon scattering region is likely to attenuate the UV radiation than that in the near-infrared, and if the hot fast-moving CO is emitted closer to the star, our view of it will be more severely affected.  \label{fig12}}
\end{figure}

\clearpage

\begin{figure}
\center
\plotone{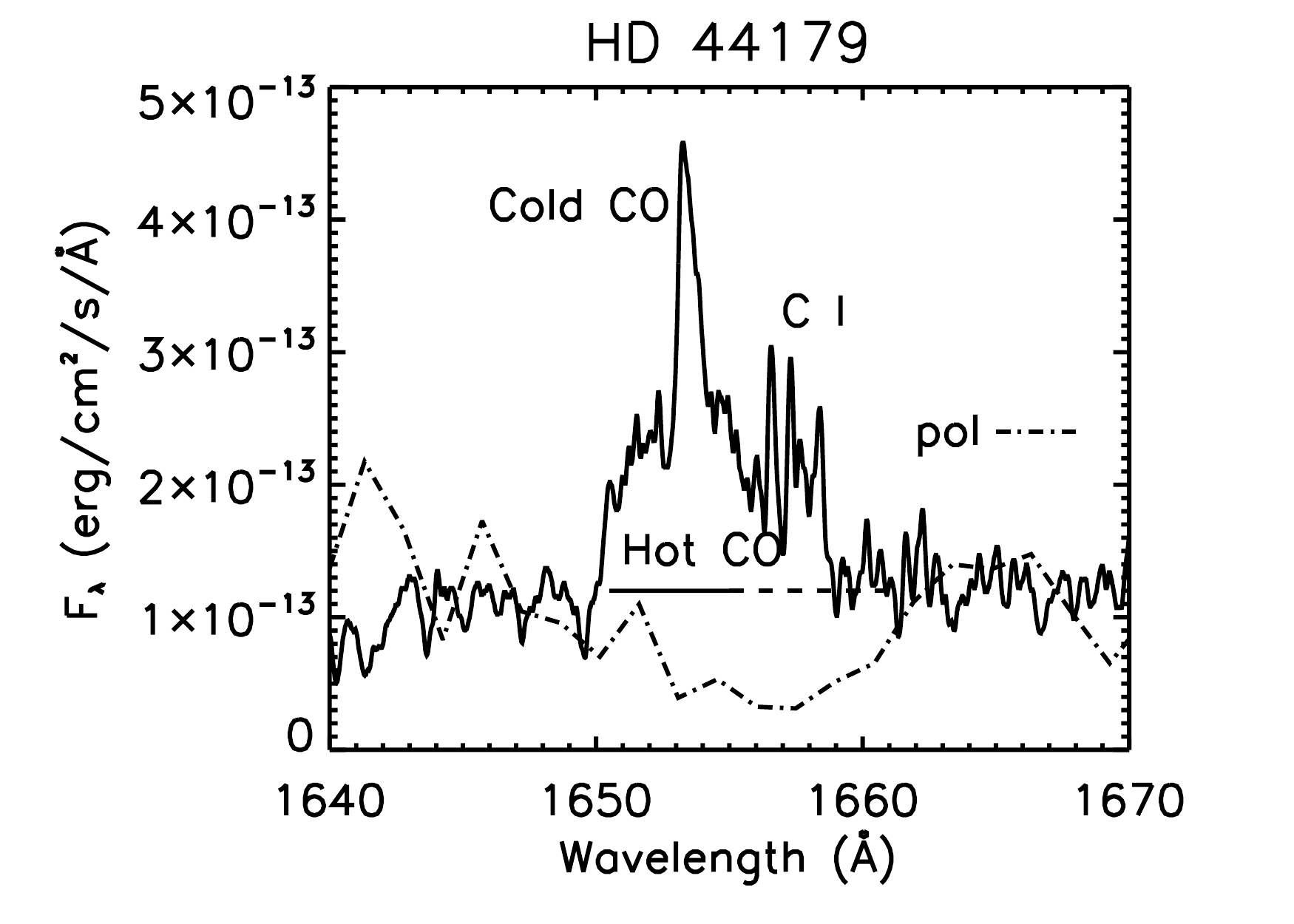}
\caption{The (0,2) A-X emission band, along with the polarization from \citep{reese96}. The polarization of the hot fast CO is greater than that of the cold slow CO, and may be the least for C I. Because the cores of the cold CO emission bands depolarize the scattered light in the system, the bulk of the cold material must be located in a region where the net scattering is less, which is naturally found further from the star. \label{fig13}}
\end{figure}

\clearpage

\begin{figure}
\center
\plotone{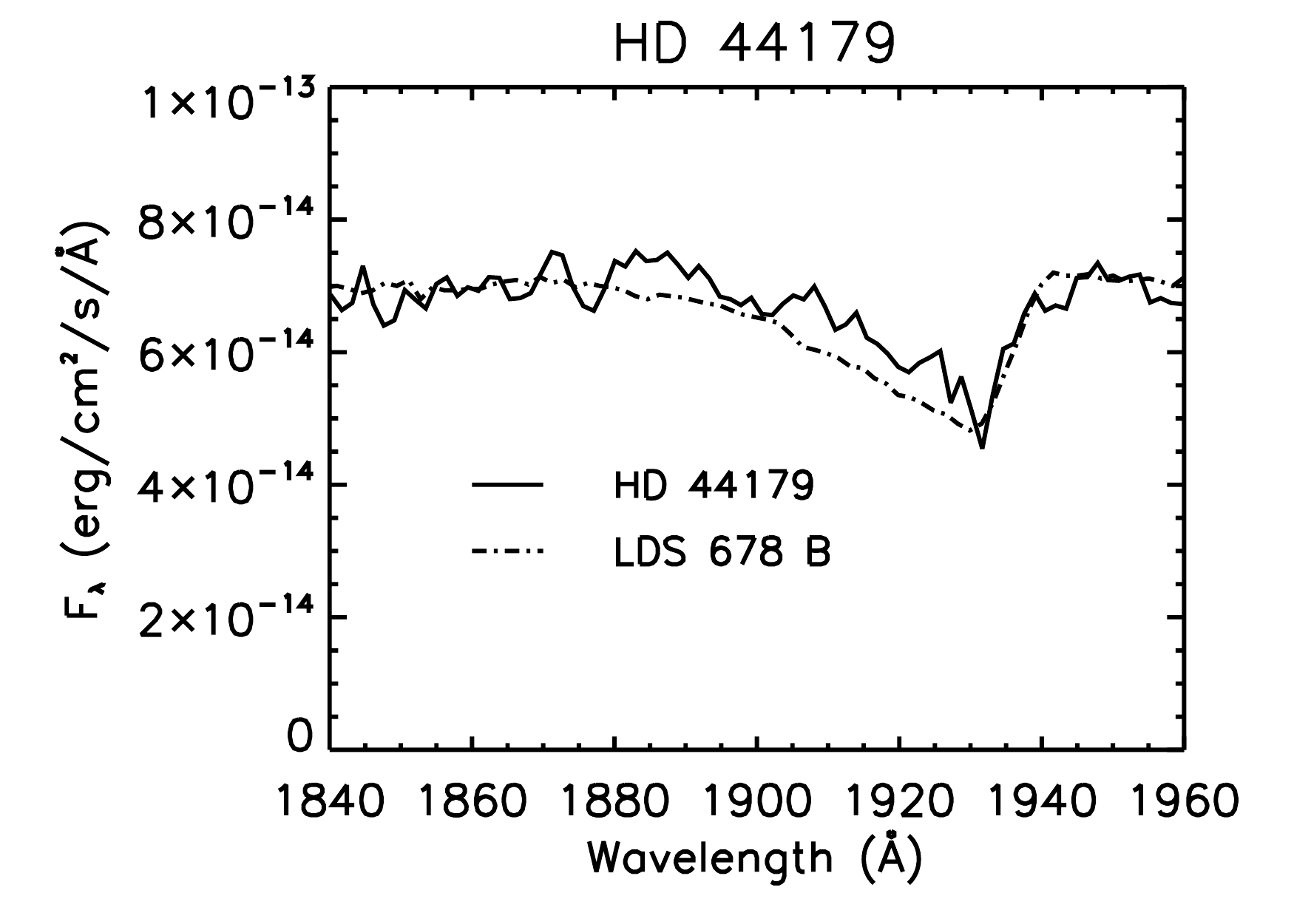}
\caption{The 1930 \AA \ feature in HD 44179, compared to the white dwarf star LDS 678 A. The latter spectrum has been diluted by scaling to 0.22 \% that  of the original, and a constant flux of  4.4x10$^{-14}$ erg cm$^{-2}$ s$^{-1}$ \AA$^{-1}$ added. Although the match to the shape of this feature is suggestive, the luminosity of such a star is not compatible with a distance to the Red Rectangle. If due to outflowing gas, the speed required for a 50 \AA \ blueshift is 8000 km s$^{-1}$. This is unlikely, especially since the hot CO gas is only traveling with a speed of 250 km s$^{-1}$. This feature may represent a complex blend of CO a-X emission and Fe II and Fe III absorption. The only other reasonable possibility is the presence of a blue-degraded molecular band that is coincident with the location of the C I line.  \label{fig14}}
\end{figure}

\clearpage

\begin{deluxetable}{lccc}
\tabletypesize{\scriptsize}
\tablecaption{Summary of Temperature and Velocity Fit Parameters \label{tbl-1}}
\tablewidth{0pt}
\tablehead{
\colhead{Component} & 
\colhead{T$_{v}$(K)}  &
\colhead{T$_{r}$(K)} &
\colhead{Velocity (km s$^{-1}$)} }  

\startdata
CO(A-X) GHRS  Emission 1580-1735 \AA & & &   \\ \hline
Very Hot Emission (0,1),(0,2),(0,3) & T$_{v^{\prime}}$=20,000 & T$_{r^{\prime}}$=5,500 & +294\tablenotemark{a}\\
Cold Emission (0,1),(0,2),(0,3) & T$_{v^{\prime}}$=2,600 & T$_{r^{\prime}}$=50 & 0\tablenotemark{b} \\
Cold Absorption (0,1) & T$_{v^{\prime\prime}}$ $\sim$500-1,000 & T$_{r^{\prime\prime}}$ $\sim$50 & 0\tablenotemark{b} \\ \hline
CO(A-X) FOS Emission 1600-2400  \AA & & & \\
(nearly identical to GHRS fit parameters) & & & \\ \hline
CO(A-X) FOS Absorption\tablenotemark{c} 1300-1600 \AA & & & \\ \hline
Hot Absorption & T$_{v^{\prime\prime}}$=2,000 & T$_{r^{\prime\prime}}$=2,000 & -294\tablenotemark{d} \\
Cold Absorption &T $_{v^{\prime\prime}}$=1,000\tablenotemark{e} & T$_{r^{\prime\prime}}$=50\tablenotemark{e} & 0\tablenotemark{d}  \\ \hline
CO(a-X) FOS Emission \\
\hline
Chemical & (approximately equal v$^{\prime}$ populations) & T$_{r^{\prime}}$=1000 & ?\tablenotemark{g} \\
\hline
OH(A-X) Emission 2550-3300 \AA & & &  \\ \hline
Very Hot Emission & NLTE\tablenotemark{f} & NLTE\tablenotemark{f} & ?\tablenotemark{g} \\
Cool Emission & (0,0) and (1,1) $\sim$ equal intensities\tablenotemark{h} & 250 & ?\tablenotemark{g} \\
\enddata
\tablenotetext{a}{Only well-determined shift from data}
\tablenotetext{b}{Shift below resolution of the data}
\tablenotetext{c}{Not fit, just estimated}
\tablenotetext{d}{Constrained by emission fit}
\tablenotetext{e}{Constrained by estimate for cold absorption}
\tablenotetext{f}{Constrained by model fit to UV photo-dissociation of H$_{2}$O observed for Space Shuttle}
\tablenotetext{g}{Shift not determined due to wavelength calibration issue. The absolute wavelength calibration of the FOS in polarimetry mode are not as precise as those in the standard mode. However, the \textit{relative} shift between the OH and Fe II, and between OH and CO (a-X) lines with the same grating \& exposures are $\sim$2.6$\pm$1.1 \AA \ and $\sim$4.0$\pm$1.3 \AA,  respectively, which are 2.4-$\sigma$ and 3.1-$\sigma$differences, suggesting possible differences in location of origin between OH and the other two species.}
\tablenotetext{h}{Populations determined by stellar pumping}
\end{deluxetable}


\begin{thebibliography}{}

\bibitem[Albritton(1989)]{alb89} Albritton, D. L., Private Communication, 1989
\bibitem[Anketell \& Learner(1967)]{ank67} Anketell, J.,  \& Learner, R.C.M., Proc. Roy. Soc. A, vol 301, p 355 (1967)
\bibitem[Bennett et al.(1999)]{ben99} Bennett, P.D., Harper, G.M.,  \& Brown, A. 1999, ASPCS 168, 231
\bibitem[Berne et al.(2008)]{berne08} Bern\'{e}, O., Joblin, C., Rapacioli, M., Thomas, J., Cuillandre, J.-C., \& Deville, Y. 2008, astro-ph 0801.3400
\bibitem[Bernstein et al.(2003)]{ber03} Bernstein, L. S., Chiu, Y., Gardner, J. A., Broadfoot, A. L., Lester, M. I., Tsiouris, M., Dressler, R. A., and Murad, E., 2003, J. Phys. Chem., 107, 10695
\bibitem[Budzien \& Feldman(1991)]{bud91} Budzien, S.A., \& Feldman, P.D. 1991, Icarus, 90, 308
\bibitem[Burke et al.(1996)]{burke96} Burke, M.L., Dimpfl, W.L., Sheaffer, P.M., Zittel, P.F., \& Bernstein, L.S. 1996, J. Phys. Chem., 100, 138
\bibitem[Cohen et al.(1975)]{cohen75} Cohen, M., et al. 1975, \apj, 196, 179
\bibitem[Cohen et al.(2004)]{cohen04} Cohen, M., H. Van Winckell, H., Bond, H.E., \& T. R. Gull,  T.R. 2004, \aj, 127, 2362
\bibitem[DeLeon(1988)]{del88} DeLeon, R. L., 1988, \textit{J. Chem. Phys.}, 89, 20
\bibitem[d'Hendecourt et al.(1986)]{dh86} d'Hendecourt, L.B., L\'{e}ger, A., Olofsson, G., \& Schmidt, W. 1986, \aap, 170, 91
\bibitem[Dieke \& Crosswhite(1962)]{die62} Dieke, G. H., \& Crosswite, H. M. 1962, J. Quant. Spectrosc. Radiat. Transfer, 2, 97
\bibitem[Di Folco et al.(2004)]{df04} Di Folco, E., Th\'{e}venin, F., Kervella, P., Domiciano de Souza, A., Coud\'{e} du Foresto, V., S\'{e}gransan, D., \& Morel, P. 2006, \aap, 426, 601
\bibitem[Dimpfl et al.(2005)]{dimpfl05} Dimpfl, W.L., Light, G.C., \& Bernstein, L.S. 2005, J. Spacecraft and Rockets, 42, 352
\bibitem[Field(1983)]{fie83} Field, R. W., 1983, J. Chem. Phys., 78, 2838
\bibitem[Glinski et al.(1996)]{glinski96} Glinski, R.J., Nuth, J.A., III, Reese, M.D.,  \& Sitko, M.L. 1996, \apj, 467, L109
\bibitem[Glinski et al.(1997)]{glinski97} Glinski, R.J., Lauroesch, J.T., Reese, M.D.,  \& Sitko, M.L. 1997, \apj, 490, 826
\bibitem[Goldman \& Gillis(1981)]{gol81} Goldman, A., and Gillis, J. R. 1981,  J. Quant. Spectrosc. Radiat. Transfer, 25, 111
\bibitem[Guelachvili et al.(1983)]{gu83} Guelachvili, G., de Villeneuve, D., Farrenq, R., Urban, W., and Verges, J., 1983, J. Molecular Spectroscopy, 98, 64
\bibitem[Harrington \& Dahn(1980)]{har80} Harrington, R.S., \& Dahn, C.C. 1980, \aj, 85, 454
\bibitem[Hidaka et al.(1982)]{hid82} Hidaka, Y., Takahashi, S., Kawano, H., Suga, M., and Gardiner, W. C. Jr., 1982, J. Phys. Chem., 86, 1429
\bibitem[Hobbs et al.(2004)]{hobbs04} Hobbs, L.M., Thorburn, J.A., Oka, T., Barentine, J., Snow, T.P., \& York, D.G. 2005, \apj, 615, 947
\bibitem[Hoogzaad et al.(2002)]{hoog02} Hoogzaad, S.N., Molster, F.J., Dominik, C., Waters, L.B.F.M., Barlow, M.J., \& de Koter, A. 2002, \aap, 389, 547
\bibitem[Jura et al.(1997)]{jura97} Jura, M., Turner, J., \& Balm, S.P. 1997, \apj, 474, 741
\bibitem[Kurucz(1976)]{kur76} Kurucz, R. L., ``The Fourth Positive System of Carbon Monoxide'', Smithsonian Astrophysical Observatory Special Report 374, 1976
\bibitem[Ledoux et al.(2001)]{led01} Ledoux, G., Guillios, O., Huisken, F., Kohn, B., Porterat, D., \& Reynaud, C. 2001, \aap, 377, 707
\bibitem[Levin et al.(1993)]{lev93} Levin, D. A., Laux, C. O., and Kruger, C. H., 1993, J. Quant. Spectrosc. Radiat. Transfer, 61, 377
\bibitem[Ma et al.(1998)]{ma98} Ma, T., Xu, J., Chen, K., Du, J., Li, W., \& Huang, X. 1998, Apl. Phys. Let., 72, 13
\bibitem[Men'shchikov et al.(2002)]{men02} Men'shchikov, A.B., Schertl, D., Tuthill, P.G., Weigelt, G., \& Yungelson, L.R. 2002, \aap, 393, 867
\bibitem[Nayfeh et al.(2005)]{nay05} Nayfeh, M.H., Habbal, S.R., \& Rao, S. 2005, \apj, 621, L121
\bibitem[Oswalt et al.(1991)]{osw91} Oswalt, T.D., Sion, E.M., Hannond, G., Vauclair, G., Liebert, J.W., Wegner, G., Koestler, D., \& Marcum, P.M. 1001, \aj, 101, 583
\bibitem[Reese \& Sitko(1996)]{reese96} Reese, M.D.,  \& Sitko, M.L. 1996, \apj, 467, L105
\bibitem[Rhee et al.(2007)]{rhee07} Rhee, Y.M., Lee, T.J., Gudipati, M.S., Allamandola, L.J., \& Head-Gordon, M. 2007, PNAS, 104, 5274
\bibitem[Rostas et al.(2000)]{ros00} Rostas, F., Eidelsberg, M., Jolly, M., Lemaire, J. L., Le Floch, A., and Rostas, J., 2000, J. Chem. Phys., 112, 4591
\bibitem[Rouse \& Engleman(1973)]{rou73} Rouse, P. E., \& Engleman, R. Jr., 1973, J. Quant. Spectrosc. Radiat. Transfer, 13, 1503
\bibitem[Saavik Ford et al.(2001)]{sf01} Saavik Ford, K.E., \& Neufeld, D.A. 2001, \apj, 557, L113
\bibitem[Sarre(2006)]{sarre06} Sarre, P.J. 2006, J. Molecular Spectroscopy, 238, 1
\bibitem[Scarrott et al.(2002)]{scarrott02} Scarrott, S.M., Watkin, S., Miles, J.R., \& Sarre, P.J. \mnras, 255, 11P
\bibitem[Schmidt et al.(1980)]{schmidt80} Schmidt, G.D., Cohen, M.,  \& Margon, B. 1980, \apj, 239, L133
\bibitem[Sitko(1983)]{sitko83} Sitko, M.L. 1983, \apj, 265, 848
\bibitem[Sitko et al.(1981)]{ssm81} Sitko, M.L., Savage, B.D., \& Meade, M.R. 1981, \apj, 246, 161
\bibitem[Soker(2005)]{sok05} Soker, N. 2005, \aj, 129, 947
\bibitem[Sundberg et al.(1993)]{sun93} Sundberg, R. L., Duff, J. W., \& Bernstein, L. S., 1993,  J. Spacecraft and Rockets, 30, 731
\bibitem[van Winckel(2003)]{vw03} van Winckel, H. 2003, \araa, 41, 391
\bibitem[van Winckel et al.(2002)]{vw02} van Winckel, H., Cohen, M., \& Gull, T.R. 2002, \aap, 147, 154
\bibitem[van Winckel et al.(1995)]{vw95} van Winckel, H., Waelkens, C., \& Waters, L.B.F.M. 1995, \aap, 293, L25.
\bibitem[Vijh et al.(2004)]{vijh04} Vijh, U.P., Witt, A.N., \& Gordon, K.D. 2004, \apj, 606, L65
\bibitem[Vijh et al.(2005)]{vijh05} Vijh, U.P., Witt, A.N., \& Gordon, K.D. 2005, \apj, 619, 368
\bibitem[Waelkens et al.(1996)]{wae96} Waelkens, C., van Winckel, H., Waters, L. B. F. M., \& Bakker, E. J. 1996, \aap, 314, L17
\bibitem[Warren-Smith et al.(1981)]{ws81} Warren-Smith, R.F., Scarrott, S.M., \& Murdin, P. 1981, Nature, 292, 317
\bibitem[Waters et al.(1998)]{waters98} Waters, L.B.F.M., Waelkens, C., van Winckel, H., Molster, F.J., Tielens, A.G.G.M., van Loon, Th., Morris, P.W., Cami, J., Bouwman, J., de Koter, A., de Jong, T., \& de Graauw, Th. 1998, Nature, 391, 868
\bibitem[Webb et al.(2001)]{webb01} Webb, R.A., Zuckerman, B., Greaves, J.S., \& Holland, W.S. 2001, AAS 197.0827
\bibitem[Wegner(1981)]{weg81} Wegner, G. 1981, \apj, 245, L27
\bibitem[Witt et al.(1998)]{witt98} Witt, A.N., Gordon, K.D., \& Furton, D.G. 1998, \apj, 501, L111
\bibitem[Witt \& Vijh (2004)]{witt04} Witt, A.N., \& Vijh, U.P. 2004, ASPCS, 309, 115.
\bibitem[Witt et al.(2006)]{witt06} Witt, A.N., Gordon, K.D., Vijh, U.P., Snell, P.H., Smith, T.L., \& Xiw, R.-H. 2006, \apj, 636, 303
\bibitem[Yan et al.(2000)]{yan00} Yan, M., Dalgarno, A., Klemperer, W.,  \& Miller, A.E.S. 2000, MNRAS, 313, L17
\bibitem[Yarkony(1992)]{yar92} Yarkony, D. R., 1992, J. Chem. Phys., 97, 1838
\bibitem[Zuckerman(2001)]{zuck01} Zuckerman, B. 2001, ARA\&A, 39, 549
\end{thebibliography}
\end{document}